%% file: main.tex
 \documentclass[preprint,12pt]{elsarticle}

\usepackage{amssymb}
\usepackage[utf8]{inputenc}
\usepackage[T1]{fontenc}

\usepackage{amsmath}
\usepackage{amsfonts}
\usepackage{mathtools}

\usepackage[ruled,vlined]{algorithm2e}

\usepackage[super]{nth}

\usepackage{dirtytalk}

\usepackage{booktabs}
\usepackage{multirow}

\usepackage{lipsum}

\usepackage{graphicx}
\graphicspath{{figure/}}
\usepackage{float}
\usepackage{subfig}

\usepackage{cleveref}
\crefname{equation}{}{}
\Crefname{equation}{Equation}{Equations}
\creflabelformat{equation}{#2(#1)#3}

\usepackage{tikz}

\include{nomenclature}

\usepackage{lineno}

\journal{International Journal of Electrical Power and Energy Systems }

\begin{document}

\begin{frontmatter}



\title{Privacy-preserving methods for smart-meter-based network simulations}

%% Alternative title
% - Smart-meters privacy-preserving methods for network simulations and energy planning
% - Privacy preserving usage of smart-meters data for network analysis
% - Two smart-meters anonymisation approaches for network simulation


\author[inst1]{Jordan Holweger \corref{cor1}}
\ead{jordan.holweger@epfl.ch}
\author[inst1]{Lionel Bloch}
\author[inst1]{Christophe Ballif}
\author[inst1]{Nicolas Wyrsch}

\affiliation[inst1]{organization={Photovoltaic and thin film electronic laboratory},%Department and Organization
            addressline={Ecole Polytechnique Fédérale de Lausanne}, 
            city={Neuchâtel},
            postcode={2000}, 
            country={Switzerland}}

\cortext[cor1]{Correspond author }

\begin{abstract}
Smart-meters are a key component of energy transition. The large amount of data collected in near real-time allows grid operators to observe and simulate network states. However, privacy-preserving rules forbid the use of such data for any  applications other than network operation and billing. Smart-meter measurements must be anonymised to transmit these sensitive data to a third party to perform network simulation and analysis. This work proposes two methods for data anonymisation that enable the use of raw active power measurements for network simulation and analysis. The first is based on an allocation of an externally sourced load database. The second consists of grouping smart-meter data with similar electric characteristics, then performing a random permutation of the network load-bus assignment.  A benchmark of these two methods highlights that both provide similar results in bus-voltage magnitude estimation concerning ground-truth voltage.  
\end{abstract}


% \begin{highlights}
% \item Two privacy-preserving methods for using smart-meter data for network planning applications are presented 
% \item The allocation method uses an external load profiles database (possibly smart-meters measurements) and the active power measurement at the transformer
% \item The smart meter anonymisation method performs a partitioning of the smart meter measurements and allows a stochastic analysis.
% \item Both methods provide accurate results with respect to transformers loading and voltage magnitude estimation
% \end{highlights}


\end{frontmatter}

%\linenumbers

%%%%%%%%%%%%%%%%%%%%%%%%%%%%%%%%%%%%%%%%%%%%%%%%%%%%%%%%%%%%%%%%%%%%%%%%%
%%%%%%%%%%%%%%%%%%%%%%%%%%%%%%%%%%%%%%%%%%%%%%%%%%%%%%%%%%%%%%%%%%%%%%%%%
%%%%%%%%%%%%%%%%%%%%%%%%%%%%%%%%%%%%%%%%%%%%%%%%%%%%%%%%%%%%%%%%%%%%%%%%%
\section{Introduction}
\label{sec:intro}

The roll-out of smart-meters (SMs) in recent decades has enabled the collection of a large amount of data. Initially, meters were designed for billing purposes \cite{Garcia2017}, but a wider range of applications is now available to utilities thanks to the valuable time-series collected. The processing of this data is often referred to as SM data analytics \cite{Wang2018}. SMs are smart only because of their bi-directional communication capabilities \cite{Zheng2013} and their ability to perform automatic reading in comparison to conventional meters \cite{Garcia2017}. This enables the shedding of particular loads by the distribution system operator (DSO). The consumer can use the (near) real-time information of the consumption data to adapt and optimize its load and reduce its energy bill. Any additional functionality comes from the processing of the collected data, regardless of whether it is performed by the SM itself or remotely. 

One well-known application is non-intrusive load monitoring (NILM) \cite{Hart1992}, which consists of disaggregating the whole-house electricity consumption into appliances/categories levels. This application is covered by extensive literature. Most applications of the NILM problem include a direct processing of the power measurement that is performed by the meter itself (or by a dedicated device) using the original high-sampling-frequency measurement \cite{Zoha2012}. Recent applications aim to disaggregate SM data in an offline phase as proposed by \cite{Holweger2019} and \cite{Zhao2020}, i.e. by using lower-sampling-rate power measurements coming from SMs (typically a sampling rate of 15 min  to 1 hour). 

The second category of SM data analytics is to build typical customer profiles and extract their characteristics. This is often referred to as load profiling \cite{Khan2018,Pilo2021}. The ultimate goal of such an identity map is to forecast flexibility as proposed by \cite{Ponocko2018a} or exploit typical load profile in a Monte-Carlo analysis \cite{Khan2017}. Those typical profiles can be used in power-system planning studies.
Alternatively, near real-time consumption information can be used to assess the network state. In an early phase of the SM roll-out, the communication infrastructure could not transmit data in real time but with a one-day delay \cite{Degefa2013}. In such a case, load forecasting was used to perform network state estimation. More recently, the time-asynchronization issue was addressed by \cite{Bahmanyar2017a} for similar purposes. The observability of the network depends on the availability \cite{Bhela2018} and location \cite{Othman2019} of the SMs. While most SMs are located in low-voltage grids, the state of the medium voltage level can be inferred from these measurements reducing the need for adding further measurement devices at the low-voltage/medium-voltage transformer \cite{Cataliotti2016,Ni2018a}. The state estimation quality may also depend on the sampling frequency, as discussed in \cite{WanYen2019}. 

All the applications mentioned above do not account for the fact that SM data contains, by nature, sensitive information such as occupancy (which can potentially be predicted  \cite{Chen2013,Becker2017,Razavi2019,Allik2020,Dorokhova2021}). For this reason, data privacy has to be tackled from a regulatory perspective \cite{Zhou2017} and from a technical perspective \cite{Sankar2013}. There are evident conflicting interests between the data owner (the households or the company being monitored) and the DSO or the energy retailer \cite{Rajagopalan2011,McKenna2012}. The researchers of \cite{McKenna2012} proposed an approach to balance both interests.  An early attempt at data anonymisation can be found in \cite{Efthymiou2010}. The researcher from \cite{Jakobi2019} acknowledges that the final application is an important consideration to whether the DSO will be allowed to use the SM data. One approach to deal with the data privacy concern is to encrypt the SM measurements to restrict the ability of a third party to use the data only for specific purposes. The review of \cite{Asghar2017} proposes an extensive overview of various data-privacy-preserving approaches to transmitting SM data. A protocol based on multiparty computations is proposed in \cite{Mustafa2019}.  A differential privacy approach is used in \cite{Gough2021}, which allows customers to weigh the importance of privacy and pay for their chosen privacy level.  Aggregation is often cited as the most computationally efficient approach to ensure privacy \cite{Eibl2021} but a drawback of aggregation is that aggregated data loses granularity and usefulness when performing the application mentioned above. Therefore, there is a need for a smart data-anonymisation approach that allows for the use of individual SM measurements in the context of power system studies. 

Our work proposes and discusses two methods to use anonymous SM data for low-voltage network simulations. We define anonymity to mean the true locations of the SMs are unknown from the third party performing the simulations. In this work, we discuss network simulations as the final applications, but our proposed methods could be applied to further types of applications such as energy planning studies. Our approach relies on the assumption that the true consumption profile of customers can be replaced by any arbitrary load profile having similar characteristics and will provide similar results when performing load-flow simulations as if using the original one. The first method consists of allocating an anonymised load profile (i.e. SM data) in a network provided that the power exchange at the medium-voltage/low-voltage transformer is measured. The second method consists of anonymising the data by grouping and permuting SM measurements. The set of possible locations for a given SM group is the only information provided to the third party. By doing so, the user has no way to retrieve the original SM locations. This paper proposes two solutions for the DSO to exploit their metering data while ensuring the anonymity of their customers.  Both approaches do not rely on advanced cryptography methods, requiring data processing at the meter level and on the third party's side. Both allow the final users to use raw SM data for their final applications.  It allows building realistic load cases for network simulations or possibly other applications like energy planning purposes. 

This paper is structured as follows. Two approaches for SM anonymisation are proposed in \cref{sec:metho}. The first is an allocation method (\cref{sec:sm_alloc}) while the second is an anonymisation by grouping and permuting method (\cref{sec:smanet}).  We compare both approaches using a dedicated benchmark presented in \cref{sec:benchmark}. \Cref{sec:results} presents key results and compares the performance, while \cref{sec:discuss} discusses advantages and drawbacks. Finally, \cref{sec:conclusion} draws some conclusions.

% To assess the impact of distributed generation on distribution grids, the following general data are required.

% \begin{itemize}
% \item grid topology
% \item building characteristics
% \item weather conditions
% \item electric and heat demands

% \end{itemize}

% Whereas the three first elements can be found for most distribution grids in Switzerland, both electric and heat demands profiles are rarely available. Regarding electricity consumption, a few distribution system operators (DSO) have already replaced conventional meters with SMs measuring the load at a resolution of 15\,min. The use of the standard demand profiles (from SIA norms) is not a viable option since these profiles aggregated at the distribution grid level result in massive electricity demand peaks due to the lack of variances. This section aims to provide solutions to cope with this issue. 

%%%%%%%%%%%%%%%%%%%%%%%%%%%%%%%%%%%%%%%%%%%%%%%%%%%%%%%%%%%%%%%%%%%%%%%%%
%%%%%%%%%%%%%%%%%%%%%%%%%%%%%%%%%%%%%%%%%%%%%%%%%%%%%%%%%%%%%%%%%%%%%%%%%
%%%%%%%%%%%%%%%%%%%%%%%%%%%%%%%%%%%%%%%%%%%%%%%%%%%%%%%%%%%%%%%%%%%%%%%%%
\section{Proposed anonymisation approaches}
\label{sec:metho}

%%%%%%%%%%%%%%%%%%%%%%%%%%%%%%%%%%%%%%%%%%%%%%%%%%%%%%%%%%%%%%%%%%%%%%%%%
%%%%%%%%%%%%%%%%%%%%%%%%%%%%%%%%%%%%%%%%%%%%%%%%%%%%%%%%%%%%%%%%%%%%%%%%%
\subsection{Smart-meter allocation}
\label{sec:sm_alloc}
The basic idea of load profile allocation is to choose, from a sufficiently large dataset of load profiles, the most-appropriate ones according to some knowledge of the network and consumers' characteristics. The first stage consists in selecting these load profiles according to their annual consumption magnitude and category (residential, commercial, etc.). The loads are allocated to network locations according to prior knowledge of the consumers' annual consumption. Each load is scaled to ensure that the total network consumption is matched. A second stage is to deform the allocated load profiles so that the resulting load at the transformer is close enough to the measured transformer load, still keeping the annual energy consumption close to the original one. 
Thus, the developed methods consist in a two-stage optimization. In the first phase, a load profile is allocated to each meter. All profiles are tuned in the second phase to match potential additional network measurements. In particular, the sum of all the profiles should be as close as possible to the profile at the transformer (which should be known).

\subsubsection{First-stage optimisation}
\begin{figure}
    \centering
    \resizebox{.6\columnwidth}{!}{\input{figure/NetworkIllust.tex}}
    \caption{Illustration of the first-stage allocation}
    \label{fig:netallocIllust}
\end{figure}
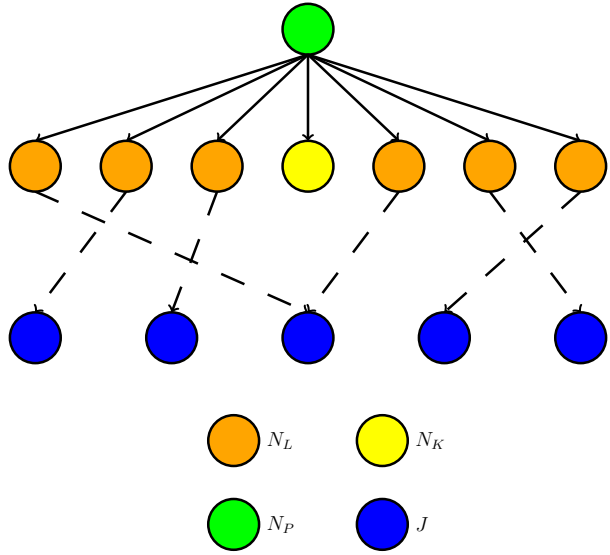

The overall idea of the first-stage optimization problem is to consider the grid as a graph, formed by a set of nodes $N$, among which the set $N_L \subset N$ of nodes has unknown load profiles. The set $N_K \subset N$ contains measured load profiles. The root node (or transformer) is denoted as $N_P \subset N$. The reference dataset of load profiles is considered as virtual nodes $J$. Any load profile is assumed to be measured on the same time span $T$. Finally, for each node $n \in N_L$ and $j \in J$ we define a load category $h_n \in H$.  The sets' definitions are given in \cref{tab:sets}. The problem can be defined as connecting each node in $N_L$ to a single node in $J$ of the same load category (as pictured in \cref{fig:netallocIllust}).   The difference between the allocated annual energy consumption of the reference load profile, $\evar_{n}$, and the one from the meter (assumed to be known for every node), $\eref_{j}$,  should be smaller than a given tolerance $\epsilon_E$. In other words, the decision variable $\beta_{n,j}$, if greater than 0, allocates and scales the load profile $j$ to the node $n$ to have the allocated annual energy $\evar_n$ close to the tolerance $\epsilon_E$ of the measured annual consumption of the node $\eref_n$. The optimization's objective is to have the minimum scaling of the available load, i.e. $\beta_{n,j} \approx 1\, \forall j \in J$. As pictured in \cref{fig:netallocIllust}, a single available load profile may be allocated to more than one node in $N_L$. A parameter of the optimization problem $k^h$  restricts the number of allocations for each load category. The size of $N_L$ may be much larger than the size of $J$. Thus, each load can be allocated more than once in the network.

\begin{table}[H]
\centering
\caption{Network topology and sets}
\begin{tabular}{@{}lll@{}}
\toprule
\textbf{Set} & \textbf{Subset of} & \textbf{Description} \\
\midrule
$N$   &    -      & network nodes\\
$H$   &  -        & load category\\
$N_P$ &  $N$      & transformer node with  measured load profiles\\
$N_L$ &  $N$      & nodes with unknown load profiles\\
$N_L^{h\in H}$ &  $N_L$  & nodes subset per load category\\
$N_K$ &  $N$      & nodes with known load profiles\\
$J$   &           & virtual nodes representing available load profiles from the dataset\\
$J^{h\in H}$ & $J$ & available load profiles subset per load category \\
$T$   &  -        & time\\

\bottomrule
\end{tabular}

\label{tab:sets}
\end{table}

The problem can be mathematically described as follows. A reference load's annual energy consumption $\eref_{j}$ is allocated to a node $n$ if the Boolean variable $\alpha_{n,j}=1$ \cref{eq:eorg}. It can be optionally scaled by a factor $\beta{n,j}>0$ \cref{eq:evar}. The scaling factor is 0 when $\alpha$ is 0 \cref{eq:alphabeta}. Only one reference load can be allocated to each node \cref{eq:oneload}. Each reference load can be allocated up to $k^h$ times \cref{eq:kalloc}.  The parameters $k^h$ are defined for each load category $h \in H$ and aim to ensure a certain variety in the choice of the allocated load. An appropriate choice for these parameters is  $k^h = \left \lceil{ \frac{N_L^h}{J^h}} \right \rceil \forall h\in H$. Assuming the annual energy demand for each node is known (either directly or by estimation), the scaled load profile should have an annual energy demand close to the reference one up to a given tolerance $\epsilon_E$ \cref{eq:e_error}. 

\begin{subequations}
\begin{align}
\label{eq:eorg}
   &\eorg_{n} = \sum_{j \in J^h} \alpha_{n,j} \eref_{j}&\quad \forall n \in N_L^h, h \in H\\
    \label{eq:evar}
    &\evar_{n} = \sum_{j \in J^h} \beta{n,j} \eref_{j}&\quad \forall n \in N_L^h, h \in H\\
    \label{eq:alphabeta}
    &\alpha_{n,j}= 
\begin{cases}
    0, & \text{if } \beta_{n,j}=0\\
    1,& \text{otherwise}
\end{cases}  &\quad \forall n \in N_L^h,j \in J^h, h\in H \\
\label{eq:oneload}
    &\sum_{j \in J^h} \alpha_{n,j} =1 &\quad \forall n \in N_L^h, h \in H\\
    \label{eq:kalloc}
    &\sum_{n \in N_L^h} \alpha_{n,j} \leq k^h &\quad \forall j \in J^h,h \in H\\
    \label{eq:e_error}
  &{\epsilon_E}^2 \geq 1 -\frac{2 \cdot  \evar_{n}}{\eref_{n}} + \frac{{ \left( \evar_{n} \right) } ^ 2 }{{ \left( \eref_{n} \right) } ^ 2 }  & \quad  \forall n\in N_L
\end{align}
\end{subequations}

The optimal allocation problem aims to make the value of $\beta_{n,j}$ as close as possible to 1. This is translated into a quadratic objective function  \cref{eq:opt_prob_alloc1} which aims to minimize the difference between the allocated energy \eorg, and the scaled energy, \evar. Besides, having the constraints applied to subset $N_L^h$ and performing a sum over $J^h$  ensure that the load category matches without introducing any additional binary variable.  All parameters and decision variables are described in \cref{tab:alloc_param}.

\begin{equation}
\label{eq:opt_prob_alloc1}
    \begin{aligned}
    \min       &\qquad&  \sum_{n \in N_L} {\left(\eorg_n \right)}^2 - 2 \cdot \eorg_n \cdot \evar_n + {\left(\evar_n\right)}^2          \\
    \text{for} & \qquad & \beta_{n,j}\\
    \text{subject to: } &\qquad&  \text{\crefrange{eq:eorg}{eq:e_error}}
    \end{aligned}
\end{equation}

%%%%%%%%%%%%%%%%%%%%%%%%%%%%%%%%%%%%%%%%%%%%%%%%%%%%%%%%%%%%%%%%%%%%%%%%%%%%%%%%%%%
\subsubsection{Second-stage optimization}
\label{sec:load_allocation_opt2}

The second stage reuses the allocated load profile, \porg \cref{eq:porg}. It tunes the allocated load profiles using a time-varying variable $\gamma_{n,t}$  \cref{eq:pvar}  to match the resulting power profile at the transformer node, $\pvar_{N_P,t}$ \cref{eq:pvarNp}  with the measured power profile $\pref_{N_P,t}$, i.e. having the relative difference between both under a given tolerance $\epsilon_P$ \cref{eq:p_error}. Additionally, the constraints on the annual energy consumption \cref{eq:evar2} still apply \cref{eq:e_error2}. The optimization problem's goal is to deform the load profiles as little as possible, hence having $\gamma_{n,t} \approx 1 \forall n \in N^L,t\in T$ \cref{eq:opt_prob_alloc2}.

\begin{subequations}
\begin{align}
\label{eq:porg}
    & \porg_{n,t} = \sum_{j \in J} \alpha_{n,j} \pref_{j,t} & \quad  \forall n \in N_L,t \in T\\
\label{eq:pvar}
   & \pvar_{n,t} = \gamma_{n,t}\porg_{n,t} & \quad  \forall n \in N_L,t \in T\\
\label{eq:pvarNp}
    & \pvar_{N_P,t} = \sum_{n\in N_L} \pvar_{n,t} + \sum_{n\in N_K} \pref_{n,t}& \quad  \forall t \in T\\
\label{eq:p_error}
    & {\epsilon_P}^2 \geq 1 -\frac{2 \cdot  \pvar_{N_P,t}}{\pref_{N_P,t}} + \frac{{ \left( \pvar_{N_P,t} \right) } ^ 2 }{{ \left( \pref_{N_P,t} \right) } ^ 2 }  & \quad  \forall t \in T\\
\label{eq:evar2}
    & \evar_n = \sum_{t\in T} \pvar_{n,t}\cdot \ts_t & \quad  \forall n \in N_L\\
\label{eq:e_error2}
     & {\epsilon_E}^2 \geq 1 -\frac{2 \cdot  \evar_{n}}{\eref_{n}} + \frac{{ \left( \evar_{n} \right) } ^ 2 }{{ \left( \eref_{n} \right) } ^ 2 }  & \quad  \forall n\in N_L 
\end{align}
\end{subequations}

\begin{equation}
\label{eq:opt_prob_alloc2}
    \begin{aligned}
    \min       &\qquad&   \sum_{n \in N_L}\sum_{t \in T} { \left( \porg_{n,t} \right) } ^ 2 -2 \cdot  \porg_{n,t} \pvar_{n,t} + { \left( \pvar_{n,t} \right) } ^ 2    \\
    \text{for} & \qquad & \gamma_{n,t}\\
    \text{subject to: } &\qquad&  \text{\crefrange{eq:porg}{eq:e_error2}}
    \end{aligned}
\end{equation}

\newcommand{\fs}[1]{\footnotesize{#1}}
\begin{table}[H]
\centering
\caption{\label{tab:alloc_param} Parameters and variables. Column \textbf{S} indicates first-stage or second-stage optimization variables.}
\begin{tabular}{@{}lccclcl@{}}
\toprule
& & \textbf{S} & \textbf{Set}  & \textbf{Dimension} & \textbf{Unit} & \textbf{Description} \\
\midrule
\parbox[t]{2mm}{\multirow{7}{*}{\rotatebox[origin=c]{90}{\textsc{parameters}}}} & 
$\pref_{n,t}$ & & $\mathbb{R}_+ $ & $ (N_P \cup N_K)\times T$  & W & measured load \\  
&$\eref_{n}$ & &$\mathbb{R}_+ $ & $N_L\cup N_K$ & J & measured annual consumption \\
&$h_n$  & & $H$ & $ N_L\cup J $ & - & load category\\
&$k^h$  & & $\mathbb{N}_+ $ & $H$ & -  & maximum allocation per category\\
&$\epsilon_E$ & & $\mathbb{R}_+ $ & &  -  & relative tolerance on energy \\
&$\epsilon_P$ & & $\mathbb{R}_+ $ & & -   & relative tolerance on power \\
&$\ts_t$ & & $\mathbb{R}_+ $ & $ T$ & s &  timesteps\\
\midrule
\parbox[t]{2mm}{\multirow{7}{*}{\rotatebox[origin=c]{90}{\textsc{variables}}}} &
$\eorg_n$ 	&\fs{1}& $\mathbb{R}_+ $ & $  N_L$ & J & original annual consumption \\
&$\evar_n$ 	&\fs{1,2}& $\mathbb{R}_+ $ & $  N_L$ & J & scaled annual consumption \\
&$\beta_{n,j}$  	&\fs{1}& $\mathbb{R}_+ $ & $N_L\times J$ & - & annual scale\\
&$\alpha_{n,j}$ 	&\fs{1}& $[0,1]$ & $N_L\times J$ &  - & allocation variable\\
&$\gamma_{n,t}$ &\fs{2}& $\mathbb{R}_+ $ & $ N_L \times T$ & - & timestep scale\\
&$\porg_{n,t}$ &\fs{2}& $\mathbb{R}_+ $ & $ ( N_L \cup N_P) \times  T$ & W & originally allocated load profiles \\
&$\pvar_{n,t}$ &\fs{2}& $\mathbb{R}_+ $ & $  ( N_L \cup N_P) \times  T$ & W & allocated load profiles\\
\bottomrule
\end{tabular}
\end{table}

%%%%%%%%%%%%%%%%%%%%%%%%%%%%%%%%%%%%%%%%%%%%%%%%%%%%%%%%%%%%%%%%%%%%%%%%%
%%%%%%%%%%%%%%%%%%%%%%%%%%%%%%%%%%%%%%%%%%%%%%%%%%%%%%%%%%%%%%%%%%%%%%%%%
\subsection{Smart-meter anonymisation}
\label{sec:smanet}

% An alternative approach to providing reliable cases study for grid planners is to use the actual SM data. The roll-out of SMs in modern distribution networks offers opportunities for large dataset acquisition. However, to use SM data, the DSO's customers must give explicit consent, as stated by the Swiss data privacy law \footnote{Federal Act
% on Data Protection (Status as of 1 March 2019), art.4 al 5}: 

% \say{If the consent of the data subject is required for the processing of personal data, such consent is valid only if given voluntarily on the provision of adequate information. Additionally, consent must be given expressly in the case of processing of sensitive personal data or personality profiles}

% Such explicit consent is, however, difficult to acquire. To comply with the law but still use these sensitive data, anonymisation shall be guaranteed. In other words, the link between the data owner and the data shall be unequivocally broken, such as one should not be able to retrieve the original data owner from the processing of its data. 

% In our case, a SM, being identified by a unique identifier, is linked to a customer by its location (either a spot in the network or a physical building address). The link between the SM identifier, building address, and the network location (referred to as ''a bus'' in the following) is critical to perform network analysis or energy system optimization as presented in this thesis. 

\begin{figure}[H]
    \centering
    \includegraphics[width=.8\columnwidth]{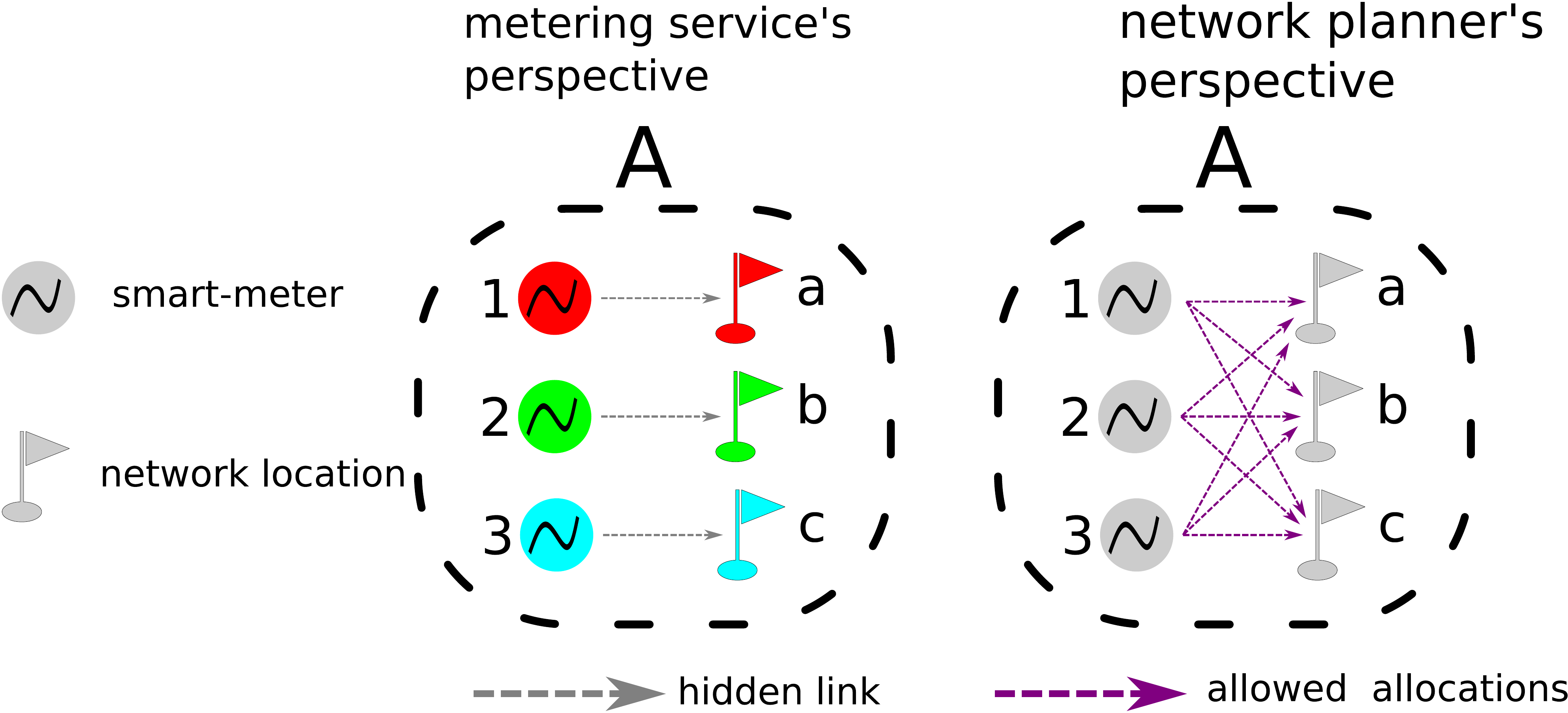}
    \caption{Graphical description of the \smanet method}
    \label{fig:smanet_schema}
\end{figure}

In this section, we propose a SM anonymisation method for network simulations (\smanet). The approach considers that the link between the SM ID ($i$) and its network location ($b$) is known from the DSO metering service. Still, it cannot be communicated to the planners or any other third party for analysis without the data owner's explicit and informed consent. However, the raw SM measurements and network topology are available for the DSO network planning service. The basic idea of this method is to group SM measurements according to some characteristic features and provide, for each group, the network location list corresponding to the group. In the graphical example of  \cref{fig:smanet_schema}, the meter IDs 1, 2, and 3 are assigned to group A (there can be as many groups as necessary). They are located in network locations a, b, and c, respectively. From the network planner's perspective, the only information accessible is that $i=$ 1, 2, and 3 are in the same group as $b=$ a, b, and c. The network planner can arbitrarily choose to allocate the SM measurements 1 to the location a, b, or c, etc. 
The underlying assumption is that the SM measurements 1, 2, and 3 are electrically similar because they are in the same group. Hence inverting $i=$ 1 and 2 at $b=$ a should have a minor impact on any further analysis performed by the network planner.

\begin{figure}[H]
    \centering
    \includegraphics[width=\columnwidth]{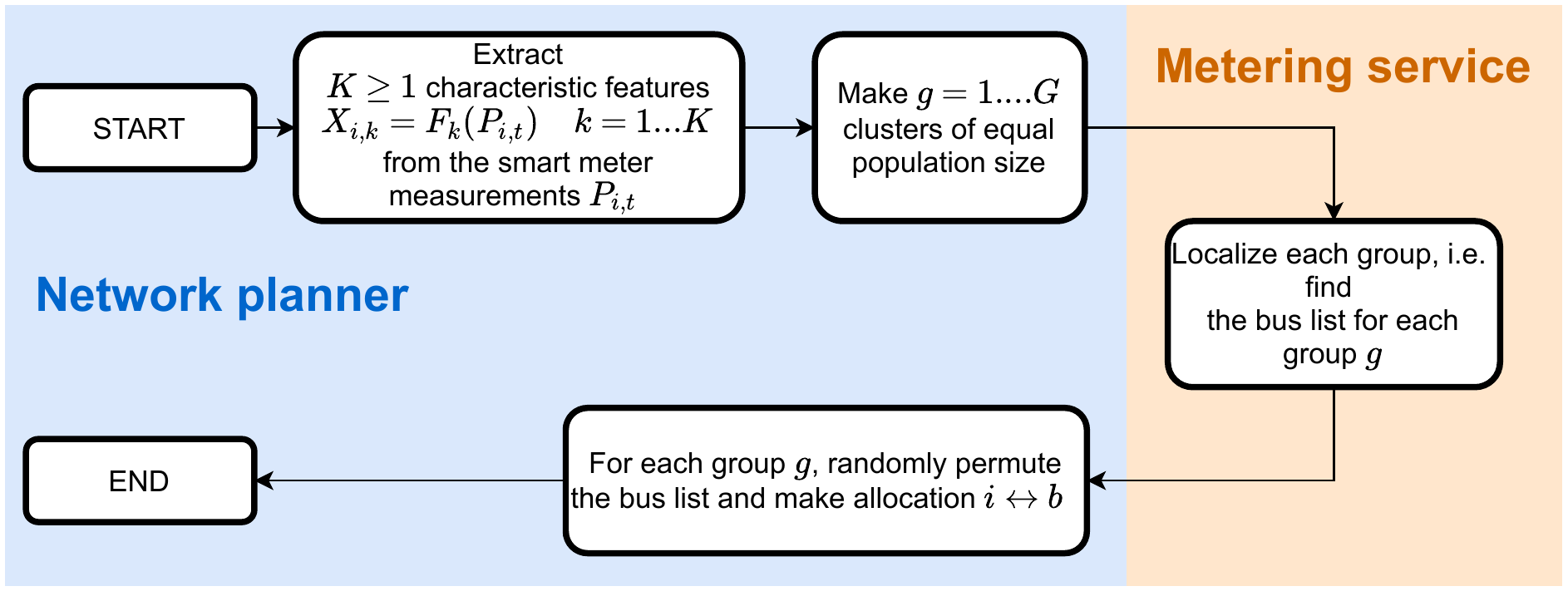}
    \caption{Workflow of the \smanet method}
    \label{fig:smanet_wf}
\end{figure}

The workflow for setting up an anonymous allocation of SM data into a designated network follows. The network planner receives the SM measurements $P_{i,t}$. The first task is to extract $K\geq1$ relevant features $X_{i,k}\,k=1...K$ for each measured load. The second task is to group the SMs according to their features. The grouping, more commonly known as clustering, has one additional constraint compared with standard clustering methods. The number of clusters is not known in advance, and the population size inside a cluster is pre-determined. Such a task is referred to as partitioning in the literature\cite{Schulz2015}. The population in each cluster should be more or less equal (balanced partitioning). At this stage, the metering service takes over and provides the list of buses for each meter group. Finally, the network planner randomly takes one permutation of the bus list to allocate each meter to a bus. This process is graphically pictured in \cref{fig:smanet_wf}.

The critical step in this process is to form groups of an equal number of elements. This might be tackled from a clustering perspective (considering that the number of elements in each cluster $m$ is predefined), such as in \cite{Chakraborty2019}. In the later derivation of the \textit{k-mean} algorithm, $m$ is quite large. For $m=2$, mathematicians consider the stable roommates' problem  \cite{Prosser2014}. The extension to triple roommates \cite{Iwama2007} or multi-dimensional roommates \cite{Lichtman2015} is still an open research topic. In the following, we investigate partitioning techniques and propose a suitable algorithm to perform the partitioning of the SM measurements into groups of equal size. 

\subsubsection{Integer programming formulation of the partitioning problem}
A generic formulation of the partitioning problem can be formulated using integer programming (IP)\cite{Miyauchi2018}. For a given dataset $\mathbb{D}$ of $N$ records and the similarity matrix $D [N\times N]$. $D_{u,v} \in [0,\infty)$ is a measure of the similarity between records $u$ and $v$.  We assume that this similarity measure respects the identity of indiscernibles ($D_{u,u}=0$),  is symmetric ($D_{u,v} = D_{v,u}$) and respects the triangle inequality ($D_{u,v} + D_{v,w} \geq D_{u,w}$). Typically such similarity metrics can be the Euclidian distance. Let $x_u$ be a vector of $K$ characteristic features. The similarity between two records can be calculated as $D_{u,v} = \sqrt{\sum_{k=1}^K \left(x_{v,k} - x_{u,k}\right)^2}$. In the following, we will consider such a similarity measure but any other similarity measure respecting the space metric properties is suitable. 

The balanced partitioning problem consists in splitting the dataset into $N_C$ clusters in which the number of records per cluster is equal for all clusters. One can deduce the prior relationship between the number of records per cluster and the number of clusters as 
\begin{equation*}
    m \leq \frac{N}{N_C} < m+1 \Rightarrow m = \left \lfloor{ \frac{N}{N_C}} \right \rfloor
\end{equation*}
\noindent where $\left \lfloor{.}\right \rfloor$ is the floor function.

Let us now see the dataset as a graph $\mathbb{G}$ where $\mathbb{S}$ is the set of edges that connect pairs of records $(u,v)$. To keep the full generality, let's assume that the graph is coarse, i.e. some pairs $(u,v)$ are not connected ($(u,v), (v,u) \notin \mathbb{S}$. The edge weights are given by the similarity matrix $D$. The partitioning problem can be seen as connecting the records or nodes of $\mathbb{G}$ to form an independent subgraph of $\mathbb{G}$ containing between $m$ and $m+1$ records. 
The selection of an edge between two records is represented by variable $\delta_{u,v}$:
\begin{equation}
     \delta_{u,v} = \begin{cases}
    1 & \text{if record $u$ is in the same cluster as $v$}\\
    0 & \text{otherwise}
\end{cases}
\end{equation}
The variable $g_{c,u}$ keeps track of the belonging of record $u$ to cluster $c$ as  
\begin{equation}
 g_{c,u} = \begin{cases}
    1 & \text{if record $u$ belongs to cluster $c$}\\
    0 & \text{otherwise}
\end{cases}
\end{equation}

Using these definitions, we can formulate the balanced partitioning problem as minimizing the sum of the selected edges' weights \cref{eq:opt_prob_mipBGP}, subject to the following constraints: A pair of unconnected nodes cannot be in the same cluster \cref{eq:ip_conn}, each node must belong to exactly one cluster \cref{eq:uInc}, the number of elements per cluster is constrained \cref{eq:clustersize}, and nodes $u \in c \text{ and } v\in c$ imply $\delta_{u,v} =1$ \cref{eq:conn1}. Conversely, $u \in c \text{ and } v\notin c$ imply $\delta_{u,v} =0$ \cref{eq:conn2}.

\begin{subequations}
\begin{align}
\label{eq:ip_conn}
 & g_{c,u} + g_{c,v} \leq 1 & \quad \forall (u,v) \notin S\\
\label{eq:uInc}
 &\sum_{c=1}^{N_C} g_{c,u} =1 &\quad \forall u=[1...N]\\
\label{eq:clustersize}  
 & m \leq \sum_{u=1}^{N_C} g_{c,u} \leq m+1 &\quad \forall c=[1...N_C]\\
\label{eq:conn1}  
 & g_{c,u} + g_{c,v} - x_{u,v}\leq 1 &\quad \forall (u,v) \in \mathbb{S}\\
\label{eq:conn2} 
 &  g_{c,u} + (1-g_{c,v}) - (1-x_{u,v})\leq 1 & \quad \forall (u,v) \in \mathbb{S}
  %& \text{each node connected to one other} & \sum_{v=1}^{N_C} \delta_{u,v} + \sum_{v=1}^{N_C} \delta_{v,u} \geq 1 &\quad \forall  u=[1...N_C] \text{ if } (u,v) \in \mathbb{S}\\ %not sure this constraints is mendatory, probably implies because in computer code (u,v) ~= (v,u)
\end{align}
\end{subequations}

\begin{equation}
\label{eq:opt_prob_mipBGP}
    \begin{aligned}
    \min       &\qquad&   \sum_{(u,v) \in \mathbb{S}} \delta_{u,v} \cdot D_{u,v}\\
    \text{for} & \qquad & \delta_{u,v}\\
    \text{subject to: } &\qquad&  \text{\crefrange{eq:ip_conn}{eq:conn2}}
    \end{aligned}
\end{equation}

\begin{figure}[H]
    \subfloat[\label{fig:ip_part_grah}llustration of graph partitioning: the red edges form the final five clusters]{\includegraphics[width=.48\columnwidth]{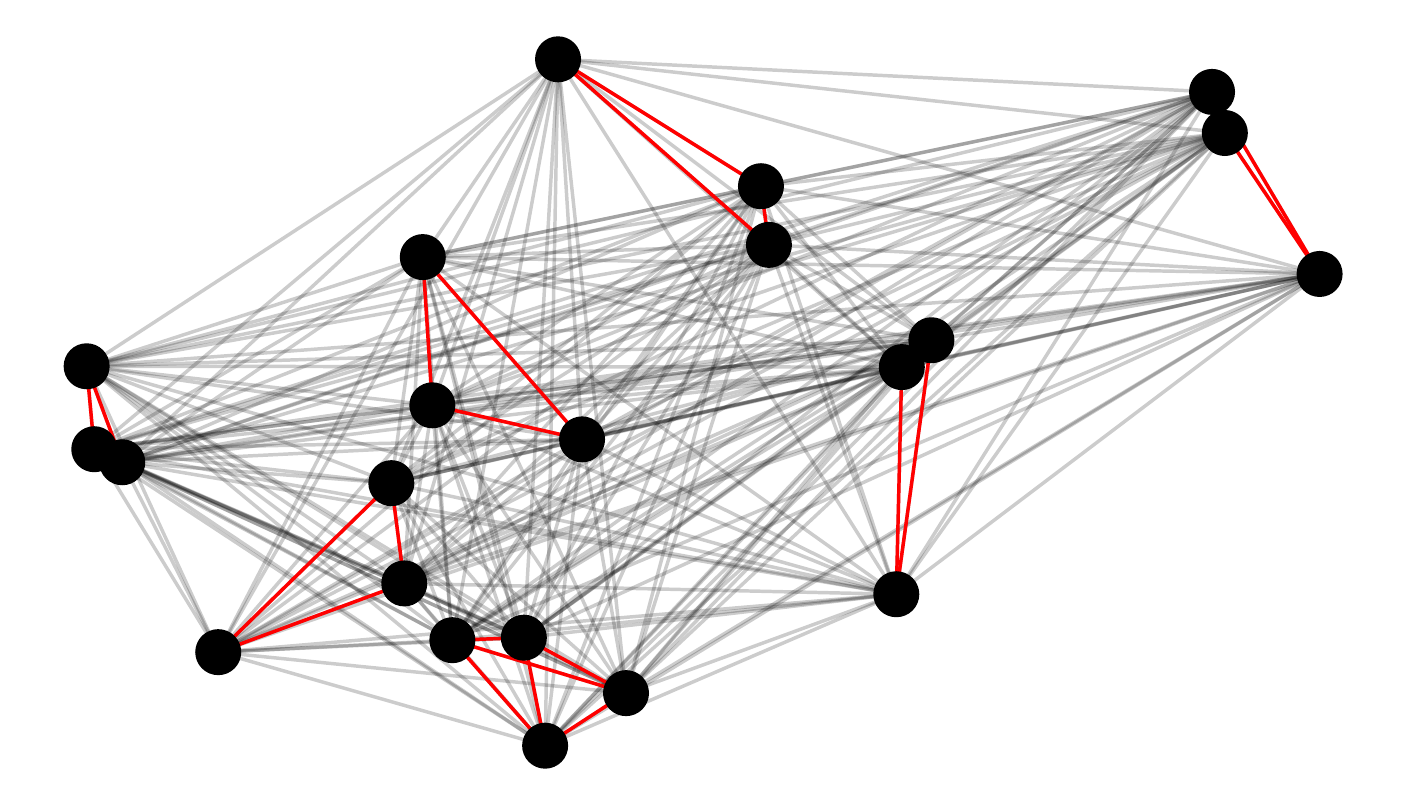}}
    \hfill
    \subfloat[\label{fig:ip_part_time}Solving time ]{\includegraphics[width=.48\columnwidth]{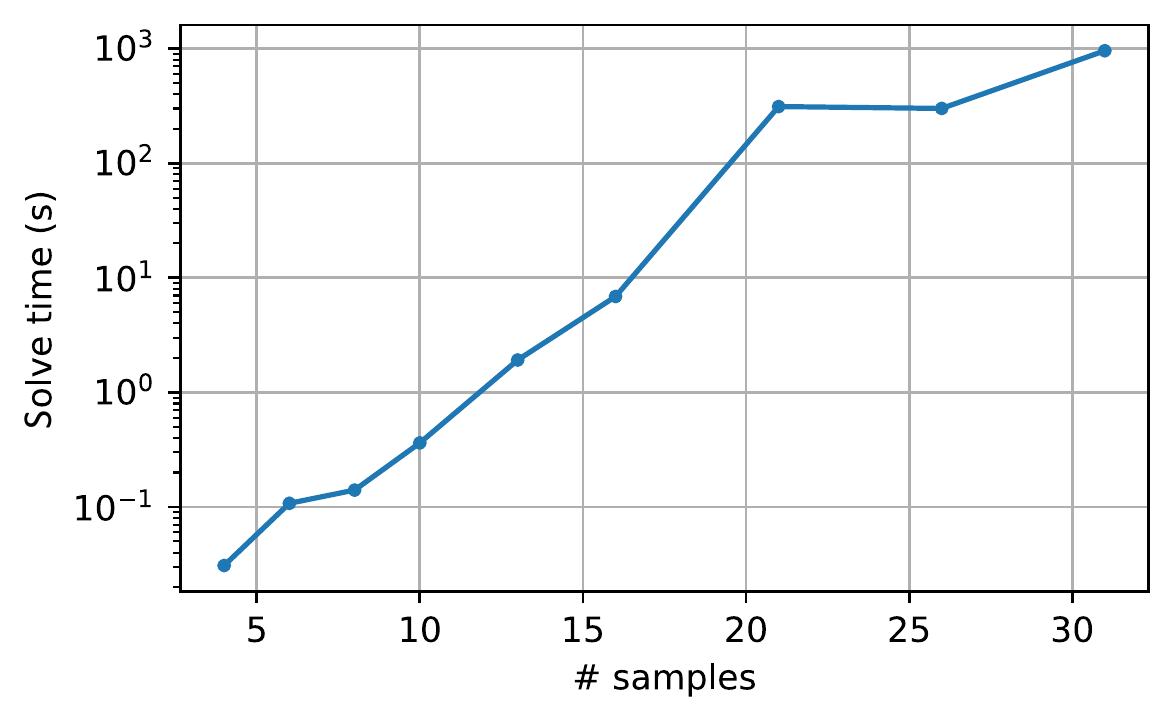}}
    \caption{Graph partitioning with integer programming}
\end{figure}

\Cref{fig:ip_part_grah}  illustrates the results of partitioning a fully connected graph of 22 nodes into seven clusters of three records. The partitioning into three records using this formulation is replicated for graphs with sizes ranging from 4 to 31 elements. For each problem, the time for solving is recorded and pictured in \cref{fig:ip_part_time}. This illustrates the issue with such a formulation. The computation time increases exponentially with the graph size. One can estimate that solving problems containing about 100 records would be in the range of $10^{8}$ years. There is hence a need for a faster partitioning method. 

\subsubsection{Spectral graph partitioning}
In modern computational science, graph partitioning is used mainly for balancing loads and minimizing scientific computation time \cite{Schulz2015} (for instance, to solve a sizeable computational fluid dynamic problem in parallel, the discretised space domain is split into smaller pieces to be individually solved on several cores).  Another application concerns route planning \cite{Schulz2015}.

Spectral graph partitioning is precisely described in \cite{Schulz2015} as the connection between cuts in a graph and its second smallest eigenvalue. To understand this relation, it is necessary to recall a few properties of a graph. Let $G$ be a graph of $n$ nodes and  $(i,j) \in S$, its set of edges. Three matrices are associated with such a graph. First, the adjacency matrix $A$ of a weighted graph is defined as 
\begin{equation*}
    A_{i,j} = \begin{cases}
        0 & \text{if } i=j\\
        w_{i,j} & \text{otherwise}
    \end{cases}
\end{equation*}

Assuming the edge weight $w_{i,j}$ represents some sort of distance between two nodes, the adjacency matrix refers to the similarity matrix presented above. Second, the degree matrix is a diagonal matrix, where each element on its diagonal is the number of edges connecting this particular node: 
\begin{equation*}
    D_{i,j} = \begin{cases}
        deg(i) & \text{if } i=j\\
        0 & \text{otherwise}
    \end{cases}
\end{equation*}
Note that $deg(i) = \sum_{(i,j)\in S} \delta_{i,j} = \sum_j \delta_{i,j}$ With: $\delta_{i,j} = 1$ if $i$ is connected to $j$, 0 otherwise. 
Finally, the Laplacian matrix is defined as
\begin{equation*}
    L = D-A
\end{equation*}

The Laplacian matrix has a few interesting properties. It is positive semi-definite, and symmetric for an undirected graph. \\
Let's now assume we perform a cut in $G$ (represented by a vector $x$) to have two distinct graphs $G_1\subset G,\,G_2 \subset G,\,G_1 \cap G_2 =  \emptyset$. The cut vector is defined as 
\begin{equation*}
    x_i = \begin{cases}
        1 & \text{if } i \in G_1\\
        -1 & \text{if } i \in G_2\\
    \end{cases}
\end{equation*}

The quadratic form of $x^TLx$ gives
\begin{equation}
\label{eq:spectral1}
    x^TLx = \sum_i\sum_j\delta_{i,j}x_i^2 - \sum_i\sum_jw_{i,j}x_ix_j 
\end{equation}
At this stage, note that 
\begin{align*}
    x_i^2&=1\\
    x_i=x_j &\Rightarrow x_ix_j = 1 \quad (i,j) \text{ is an uncut edge}\\
    x_i=-x_j &\Rightarrow x_ix_j = -1 \quad (i,j) \text{ is a cut edge}\\
\end{align*}
We can rewrite \cref{eq:spectral1} as 
\begin{equation}
\label{eq:spectral2}
x^TLx = \underbrace{\sum_i\sum_j\delta_{i,j}}_{\text{number of edges in $G$}} - \sum_{(i,j) \text{uncut}}w_{i,j} + \sum_{(i,j) \text{cut}}w_{i,j}  
\end{equation}
Hence maximizing \cref{eq:spectral2} is equivalent to finding a cut that splits $G$ into the two most-distant parts. Due to the Laplacian matrix properties, this is equivalent to finding the highest eigenvalue and using the corresponding eigenvector to perform the cut (detailed derivations are given in \cite{Schulz2015}). 
The spectral graph partitioning algorithm can be written as

\begin{algorithm}[H]

\SetAlgoLined
\SetKwInOut{Input}{input}\SetKwInOut{Output}{output}
\Input{$G$ a weighted graph}
\Output{$G' = G_1 \cup G_2$ with $G_1\subset G,G_2 \subset G$ distinct graphs, $G_1 \cap G_2 =  \emptyset$}
\BlankLine
$L$: Laplacian of $G$\;
$v,\lambda$ eigenvectors and associated values  of $L$\;
Get the maximum eigenvalue and associated vector: $v^\tf{max},\lambda^{\tf{max}} = \max_\lambda \lambda$ \;
$m = \tf{median}(v^\tf{max})$\;
Construct cut vector $x$ as: $x_i = \begin{cases}
        1 & \text{if } v^\tf{max}_i >=m\\
        -1 & \text{if } v^\tf{max}_i <m
    \end{cases}$\;
$G_1,G_2 = \tf{cut}(G,x)$\;
$G' = G_1 \cup G_2$ \;
\Return $G'$
 \caption{\label{alg:sgp} Spectral graph partition (SGP)}
\end{algorithm}

% \begin{itemize}
%     \item Get the Laplacian matrix $L$
%     \item Compute the associated eigen value by solving $Lv=\lambda v$
%     \item take the highest eigen value $\lambda^{\tf{max}}$ and associated eigen vector $v^\tf{max}$
%     \item compute the median $m$ of the $v^\tf{max}$ components
%     \item Retrieve the cut vector as $x_i = \begin{cases}
%         1 & \text{if } v^\tf{max}_i >=m\\
%         -1 & \text{if } v^\tf{max}_i <m
%     \end{cases}$
% \end{itemize}

\begin{figure}[H]
    \centering
    \includegraphics[width=.6\columnwidth]{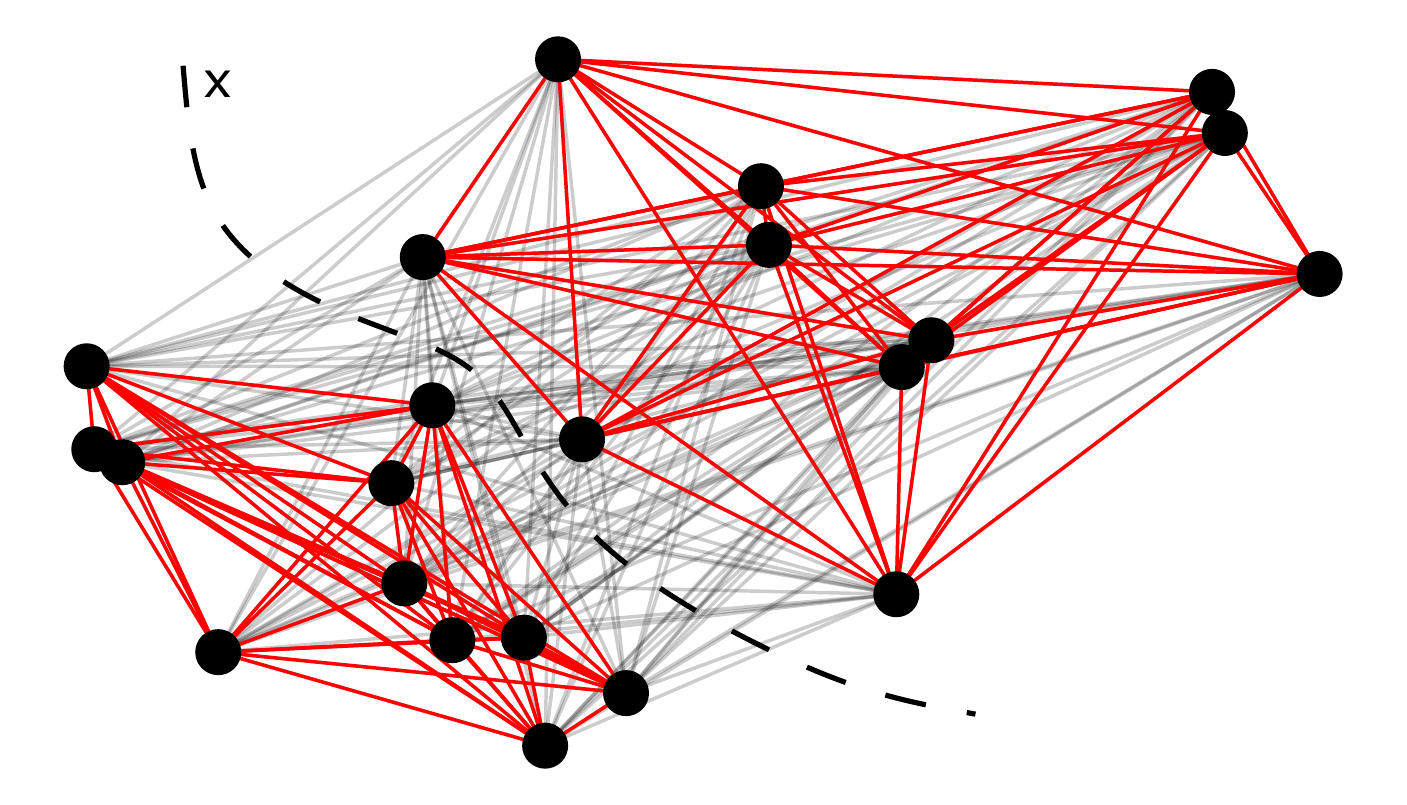}
    \caption{Illustration of a single cut in a random graph, represented by its nodes (black dots) and edges (gray and red lines)}
    \label{fig:graph_cut}
\end{figure}

A graphical example of a single cut through a graph using the graph partitioning algorithm is pictured in \cref{fig:graph_cut}. 

\begin{figure}[H]
    \centering
    \includegraphics[width=\columnwidth]{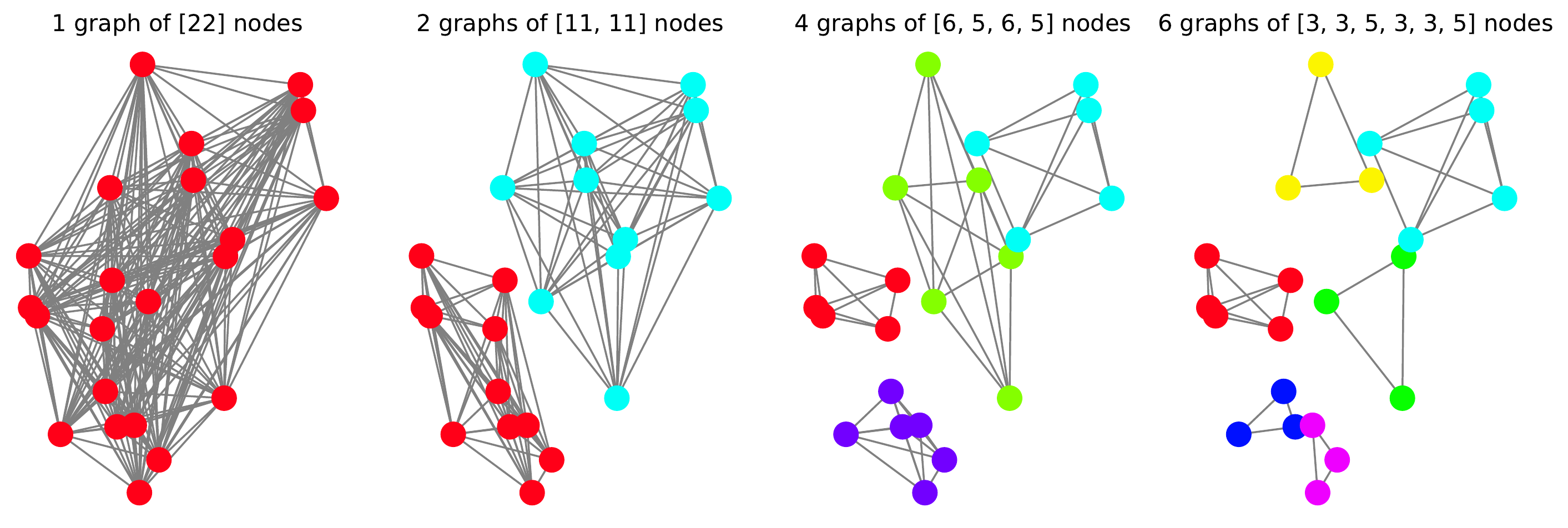}
    \caption{Illustration of successive (from left to right) spectral graph partitioning}
    \label{fig:graph_exemple}
\end{figure}

This approach can be used to successively cut the original graph into smaller partitions until the size of a sub-graph is smaller than $2k$ for $k$ the desired graph size. The resulting clusters will have a size between  $k$ and $2k-1$ as illustrated in \cref{fig:graph_exemple}. For $k>2$, this can lead to significant unbalance. An IP formulation could be used to further reduce the imbalance when the next graph to cut has a relatively small number of nodes.   

\subsubsection{Recursive spectral graph partitioning}
The spectral graph partitioning has low complexity and ensures more or less balanced clusters when applied successively. In contrast, the IP formulation of the graph partitioning problem  provides the most optimal balancing of the graph partitioning, but the computational complexity makes it unusable for large graphs. To gain the best of the two worlds, we propose the following recursive spectral graph partitioning (RSGP) algorithm:

\begin{algorithm}[H]
\label{alg:RSGP}
\SetAlgoLined
\SetKwInOut{Input}{input}\SetKwInOut{Output}{output}
\Input{$G$ a weighted graph, $k$ the desired partition size}
\Output{$G' = \bigcup_{j=1}^J {G_j}$ with $G_j$ a set of distinct graphs $G_j \subset G, G_l \cap G_k =\emptyset \quad  l,k ={[1...J] \,l\neq k}$}
\BlankLine
$[G_1,G_2] = \tf{SGP}(G)$: cut of $G$ using spectral graph partitioning (\cref{alg:sgp}) \;
$G'$: an empty graph \;
\For{i in [1,2]}{
    $N_i$ : number of nodes in $G_i$\;
    \uIf{ $3k \leq N_i < 4k$}{
        $G_i'$: apply IP formulation to partition $G_i$ \cref{eq:opt_prob_mipBGP} \;
        }
    \uElseIf{$N_i \geq2k$}{
        $G_i' = \tf{RSGP}(G_i)$\;
        }
    \Else{
        $G_i' = G_i$\;
        }
    $G' = G' \cup G_i'$\;
}
\Return $G'$
 \caption{ Recursive spectral graph partition (RSGP)}
\end{algorithm}

\begin{figure}[H]
    \subfloat[Clusters obtained with successive SGP: three clusters of three records and three clusters of five records ]{\includegraphics[width=.3\columnwidth]{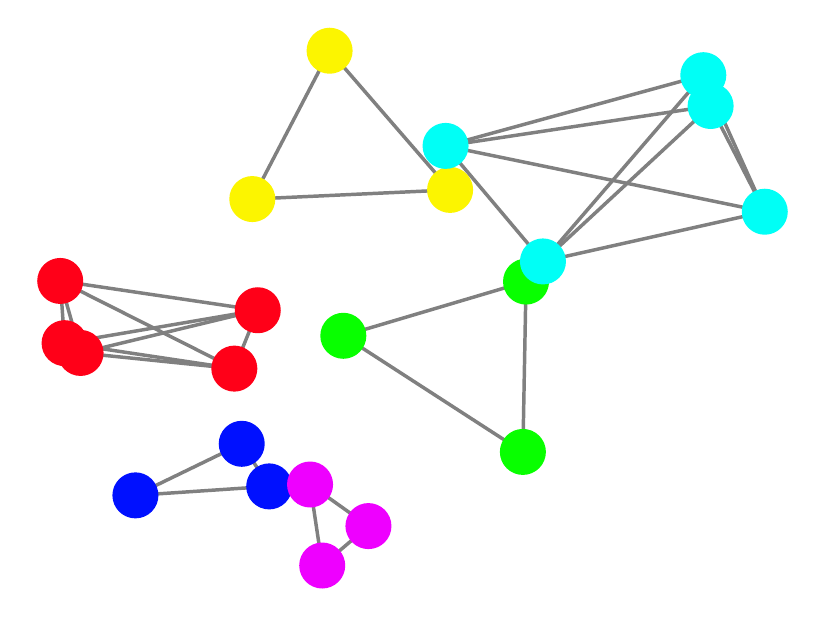}}
    \hfill
     \subfloat[Clusters obtained with RSGP: two clusters of three records and four clusters of four records  ]{\includegraphics[width=.3\columnwidth]{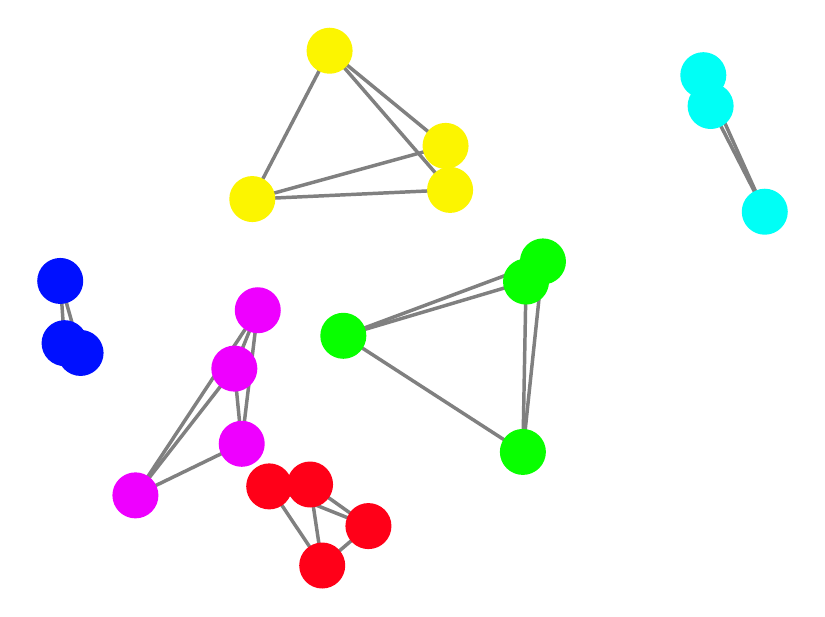}}
     \hfill
     \subfloat[Clusters obtained with IP: six clusters of three records and one cluster of four records]{\includegraphics[width=.3\columnwidth]{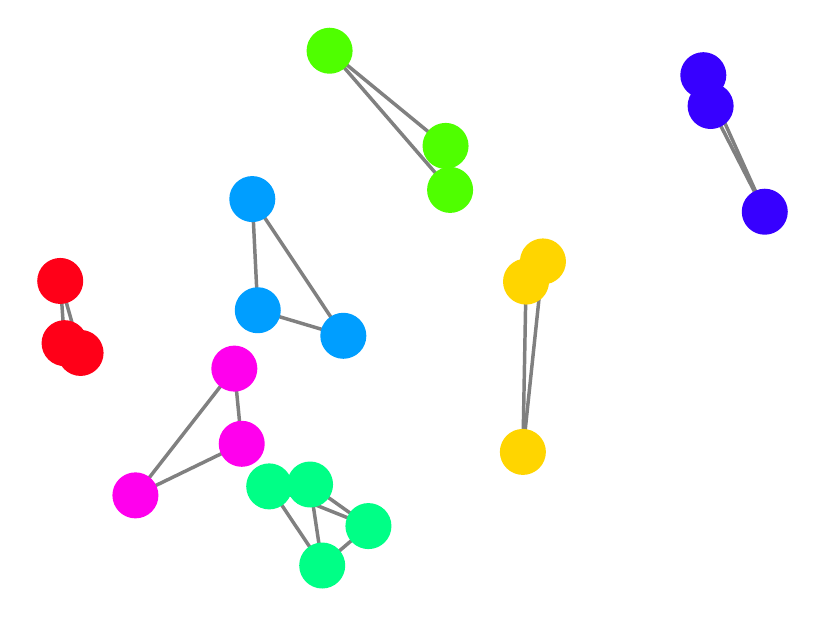}}
    \caption{Comparison of the partitioning of 22 nodes with the SGP, RSGP, and IP methods}
    \label{fig:graph_comp}
\end{figure}

\begin{figure}[H]
    \centering
    \includegraphics[width=.7\columnwidth]{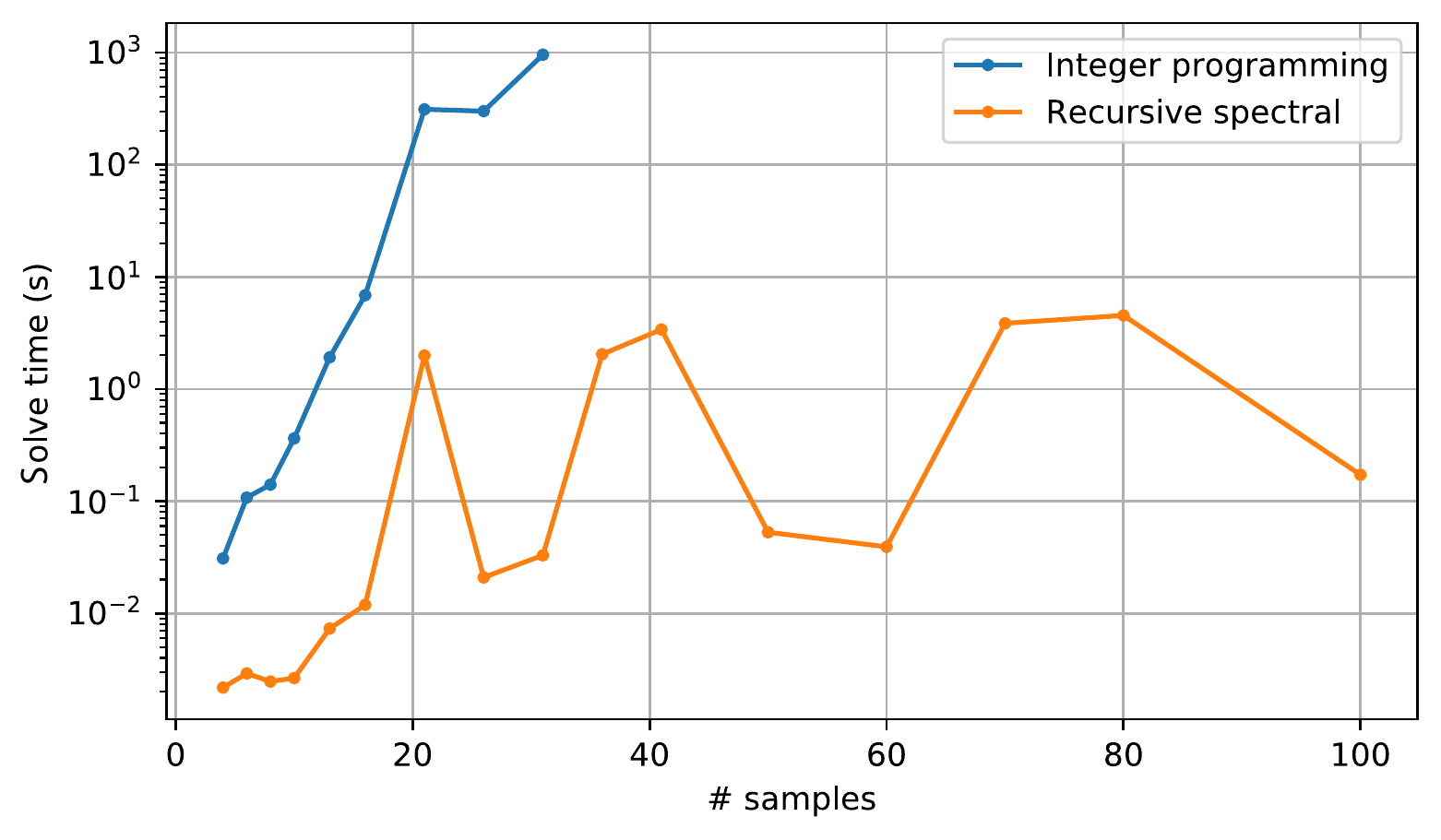}
    \caption{Computation time comparison}
    \label{fig:comp_time_comp}
\end{figure}

The comparison of the partitioning of the graph with 22 nodes into clusters of three records using  SGP, RSGP, and IP is presented in \cref{fig:graph_comp}. In terms of unbalance ($\sum_g N_g-k$, with $N_g$ the $g^\tf{th}$ partition size),  the proposed RSGP algorithm lies in between the successive SGP algorithm and the IP optimization (RSGP's unbalance is four vs. six for SGP, and one for IP). However, the RSGP computing time does not increase exponentially with the number of nodes, as shown in \cref{fig:comp_time_comp}.  

Hence, the RSGP algorithm is suitable for partitioning large datasets into small groups of equal size and fulfilling the requirements for grouping SM measurements for anonymisation purposes. So far, the definition of the SM features (the record attributes) has not been discussed. Indeed the feature selection depends on the ultimate goal of the study, as it will determine which load characteristics are significant and how similarity is defined. Provided that the selected SM features (i.e. load characteristics) best reflect the goal of the analysis, the RSGP   supposes that any intra-group permutation of one SM by another will provide similar results.   In this work, we assume that the SM data are used for ''network simulations''. For this specific end use, we will discuss the most appropriate features.

%%%%%%%%%%%%%%%%%%%%%%%%%%%%%%%%%%%%%%%%%%%%%%%%%%%%%%%%%%%%%
%%%%%%%%%%%%%%%%%%%%%%%%%%%%%%%%%%%%%%%%%%%%%%%%%%%%%%%%%%%%%
\section{Benchmark}
\label{sec:benchmark}
We have proposed two methods to enable network simulations using real load measurements: 
\begin{itemize}
    \item The load profiles allocation technique
    \item the load profiles anonymisation technique
\end{itemize}

To validate these two approaches and compare their performance with respect to network simulations, we solve the load-flow equation using reference loads. The load-flow problem allows us to calculate the buses' voltage ($V_{b,t} \quad b\in \text{Bus set}$), lines' current ($I_{l,t} \quad l \in \text{line set}$), and the power at the substation (transformer power $P^\tf{trafo}_i$) at all time $t\in T$.  The load-flow is then solved using the loads resulting either from the load allocation technique or from one random permutation resulting from the anonymisation technique. The performance of the two methods are measured using dedicated key performance indicators: 

\begin{align}
    & \text{Voltage magnitude mean squared error} & \tf{MSE}_\tf{vm} &= \sum_{t=1} ^T\sum_{b=1}^B \frac{\left(V_b -V^\tf{ref}_b\right)^2}{B} \\
    & \text{Maximum transformer loading error} & \tf{E}_\tf{maxTRL}&= \frac{P^\tf{trafo}_\tf{max} - P^\tf{trafo,ref}_\tf{max}}{P^\tf{trafo,ref}_\tf{max}} \\
    & \text{Maximum line loading error} & \tf{E}_\tf{maxLNL} &= \frac{I_\tf{max} - I^\tf{ref}_\tf{max}}{I^\tf{ref}_{l,t}  } \\
    & \text{Minimum voltage error} & \tf{E}_\tf{minVM} &= \frac{V_\tf{max} - V^\tf{ref}_\tf{min}}{V^\tf{ref}_\tf{min}}
\end{align}
\noindent where subscripts min and max denote the minimum or maximum over the time and element index ($b$ for the buses, $l$ for the lines, $i$ for the transformers), superscript ref indicates the reference case values.

\subsection{Reference case}

The DSO \textit{Romande Energie} deployed SMs in the Rolle area. Those data are not accessible for privacy reasons, as explained above. In this work, we assume that if at least three customers are metered by SMs for a given location in a network, aggregating the consumption to a single virtual SM is enough to preserve individual customers' privacy. This is typically the case for multi-family buildings. In this case, the original SM measurements have been aggregated and only the resulting consumption has been provided. Those SMs are attributed to their actual network location. If two or more meters measure the same customer's consumption (measuring different circuits), they are also considered as one single SM and aggregated together (their location in the network is not known). All other SM measurements have been allocated in the networks by matching their annual energy consumption with the estimated building consumption (using the SIA norms \cite{SIA2015} and a method inspired by \cite{girardin_gis-based_2012}). The matching basically consists of minimizing the sum of the difference between the buildings' and SMs' annual energy consumption.

\begin{figure}[H]
    \centering
    \includegraphics[width=.9 \columnwidth]{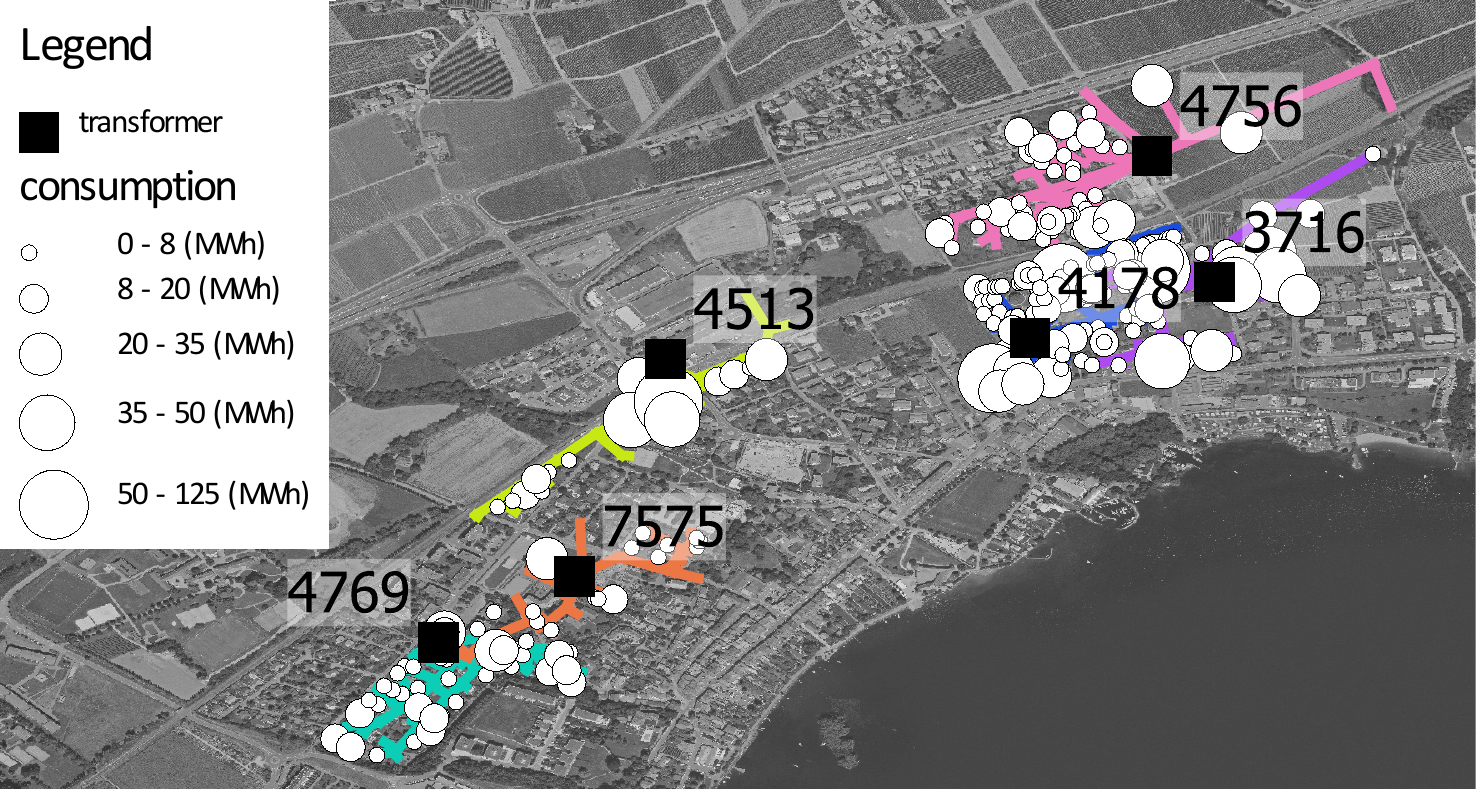}
    \caption{Map of the six networks' and reference loads' annual consumption. The numbers next to the transformers indicate the network ID.}
    \label{fig:map_six_net}
\end{figure}

Finally, SMs that appear to measure consumption and production (a \pv system in a self-consumption scheme) have been discarded. 
In the end, 257 SM measurements are used in this reference case. The annual energy consumption and network lines from six low-voltage networks are shown in \cref{fig:map_six_net}. 

\subsection{Load allocation database and parameters}
For the load allocation method, a database of SM measurements is required. The loads are split into three categories: Apartment, House, and Not residential (hereafter Not res.). The load database gathers SM measurements from a large set of (non-)residential sites. The measurements were acquired during the \flexi \cite{flexi}, and \flexii \cite{flexi2} projects. These projects concerned networks outside the Rolle area. The number of loads in the database (reported in \cref{tab:load_db}) has to be compared with the number of loads present in each network and category (\cref{tab:load_net}). The parameter $k^h$, representing the maximum number of allocations of a load in a particular network, is obtained by dividing the number of loads in the network by the number of loads in the database for a given category. All measurements have a resolution of 15\,mins and cover one year. The energy and power tolerances for the allocation ($\epsilon_E$ and $\epsilon_P$) are set to 5 and 1\%, respectively. 

\begin{table}[H]
\centering
\caption{\label{tab:load_net} Number of loads per category ("Not res." means not residential) and median annual consumption for each network (TR\#)}
\begin{tabular}{@{}llcc@{}}
\toprule
\textbf{TR \# }  & \textbf{Category}  & \textbf{\# load } & \textbf{Median cons. (MWh)}\\ 
\midrule
\parbox[t]{2mm}{\multirow{3}{*}{\rotatebox[origin=c]{90}{\textbf{3716}}}}
        & Apartment & 13       & 11.2               \\
        & House     & 15       & 5.4               \\
        & Not res.    & 22       & 11.6             \\
\midrule
\parbox[t]{2mm}{\multirow{3}{*}{\rotatebox[origin=c]{90}{\textbf{4178}}}}
        & Apartment & 9        & 20.2           \\
        & House     & 51       & 4.5            \\
        & Not res.    & 6        & 10.0         \\
\midrule
\parbox[t]{2mm}{\multirow{3}{*}{\rotatebox[origin=c]{90}{\textbf{4513}}}}
        & Apartment & 18       & 4.1              \\
        & House     & 3        & 3.7              \\
        & Not res.    & 9        & 11.0             \\
\midrule
\parbox[t]{2mm}{\multirow{3}{*}{\rotatebox[origin=c]{90}{\textbf{4756}}}}
        & Apartment & 3        & 4.3              \\
        & House     & 28       & 5.4             \\
        & Not res.    & 18       & 9.9             \\
\midrule       
\parbox[t]{2mm}{\multirow{3}{*}{\rotatebox[origin=c]{90}{\textbf{4769}}}}
        & Apartment & 4        & 4.2         \\
        & House     & 26       & 4.9           \\
        & Not res.    & 14       & 12.5           \\
\midrule
\parbox[t]{2mm}{\multirow{3}{*}{\rotatebox[origin=c]{90}{\textbf{7575}}}} 
        & Apartment & 2        & 65.9         \\
        & House     & 11       & 4.9          \\
        & Not res.    & 5        & 5.1         \\
\bottomrule
\end{tabular}
\end{table}

\begin{table}[H]
\centering
\caption{\label{tab:load_db} Number of loads per category (Not res. means not residential) and median annual consumption for each database source }
\begin{tabular}{@{}llcc@{}}
\toprule

 \textbf{Src.}  & \textbf{Category}  &   \textbf{\# load } & \textbf{Median cons. (MWh)} \\ 
\midrule
\parbox[t]{2mm}{\multirow{3}{*}{\rotatebox[origin=c]{90}{\flexi}}}
        & Apartment         & 38       & 3.3     \\
        & House             & 46       & 4.4      \\
        & Not res.          & 1        & 22.4   \\
\midrule
\parbox[t]{2mm}{\multirow{3}{*}{\rotatebox[origin=c]{90}{\flexii}}} 
        & Apartment & 44       & 2.1                 \\
        & House     & 48       & 4.2                 \\
        & Not res.  &          &                   \\
\midrule
\multicolumn{2}{c}{\textbf{Not res.}}    
           & 3        & 407                \\
\bottomrule

\end{tabular}
\end{table}

% SQL query to complete above table (replace flexi_1 by the other projects
%% FLEXI 1
% SELECT housetype,count(t3.id_profile) as nb_load, percentile_cont(0.5) within group (order by t3.annual_cons) as med_cons FROM (
% 	SELECT flexi1_quest_feat .id_profile, annual_cons, housetype FROM flexi1_quest_feat 
% 	LEFT JOIN (
% 		SELECT id_profile,sum(e_imp) as annual_cons FROM measure 
% 		WHERE time>= '2013-11-10' AND time <'2015-01-10' AND id_profile IN (
% 			SELECT id_profile FROM profile WHERE id_source='flexi_1') 
% 		GROUP BY id_profile
% 	) t2
% 	ON flexi1_quest_feat.id_profile=t2.id_profile
% ) AS t3
% GROUP BY housetype

%% FLEXI2 
% SELECT hometype,count(t3.id_profile) as nb_load, percentile_cont(0.5) within group (order by t3.annual_cons)/1000 as med_cons FROM (
% 	SELECT flexi2_quest_feat .id_profile, annual_cons, hometype FROM flexi2_quest_feat 
% 	LEFT JOIN (
% 		SELECT id_profile,sum(e_imp) as annual_cons FROM measure 
% 		WHERE time>= '2014-11-10' AND time <'2016-01-10' AND id_profile IN (
% 			SELECT id_profile FROM profile WHERE id_source='flexi_2') 
% 		GROUP BY id_profile
% 	) t2
% 	ON flexi2_quest_feat.id_profile=t2.id_profile
% ) AS t3
% GROUP BY hometype

% COMMERCIAL 

\subsection{Smart-meter anonymisation method}
\label{sec:sm_anony_method}

\begin{figure}[H]
    \centering
    \subfloat[\label{fig:e_pmx}Energy and maximum power distribution]{\includegraphics[width=.5 \columnwidth]{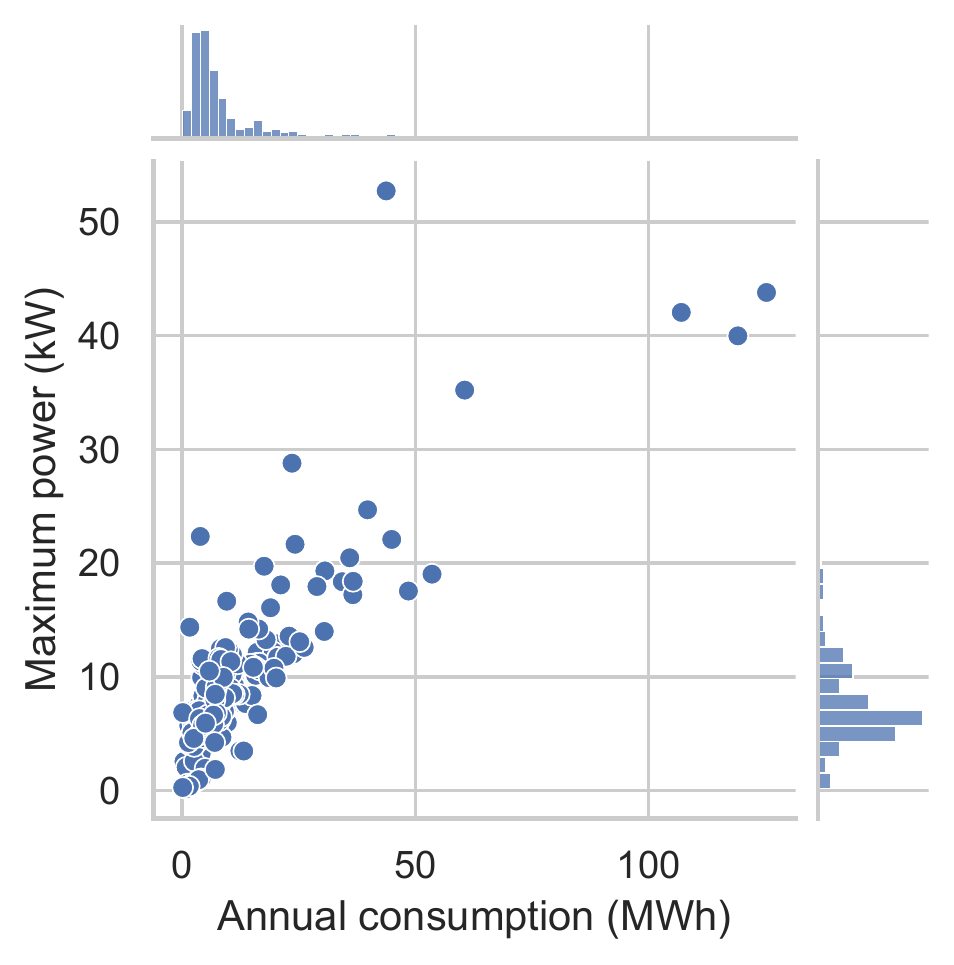}}
    \subfloat[\label{fig:clust_e_pmax1} Resulting groups colored by group ID]{\includegraphics[width=.5 \columnwidth]{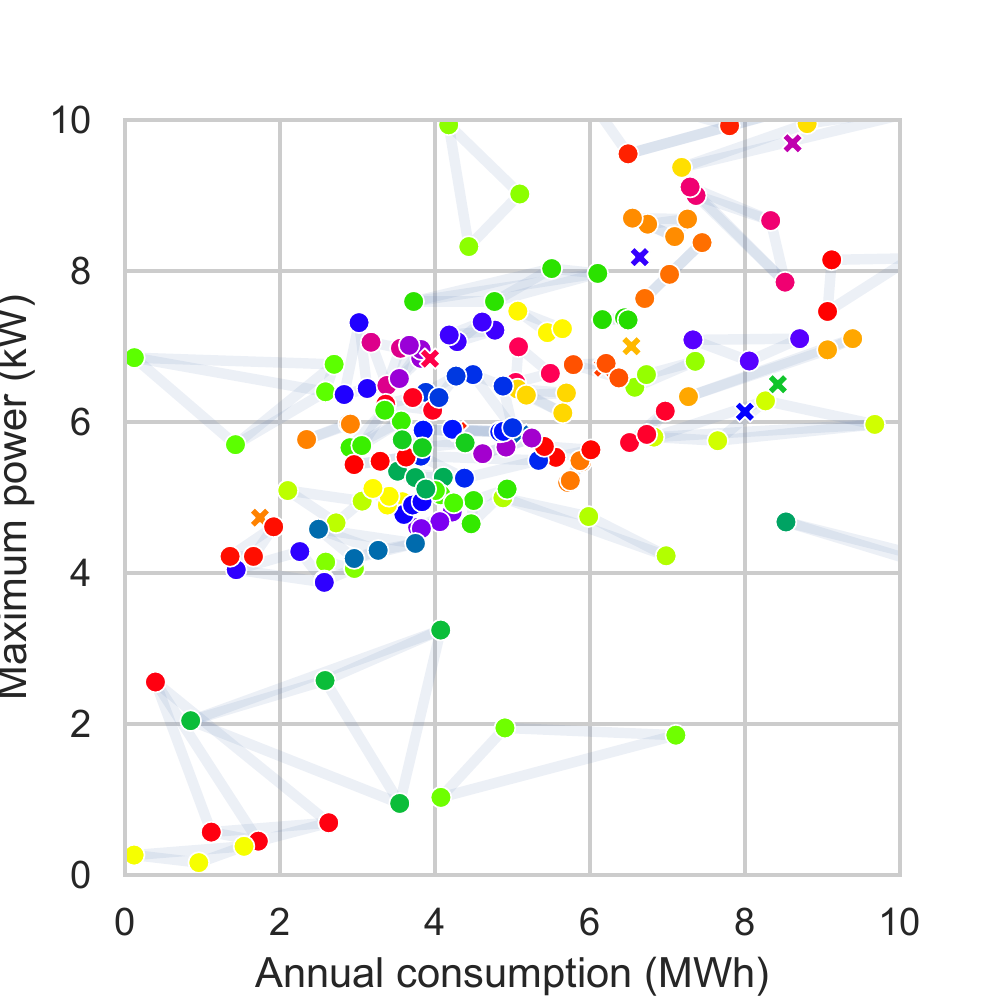}}
    \caption{SM measurements' selected features and results of the partitioning }
    \label{fig:cluster_map}
\end{figure}

As stated in the reference case description, \RE provided the 257 SM measurements located in the six sub-networks. To mimic a real case, the measurements' true locations are unknown except for the measurements grouping three or more customers (39 loads in total). For all other loads, the SM anonymisation (\smanet) technique should be applied. The first step is to define the features of the loads. In a primary approach, the energy ($E_i=\sum_t P_{i,t}\cdot \ts_t$), and maximum power ($P^\tf{max}_i = \max_t P_{i,t}$) are used as input features for the partitioning. (The features are plotted in \cref{fig:e_pmx}.) The second step is to perform the partitioning using the RSGP method. For this step, the features are normalized to have zero mean and unity variance before the distance matrix $D$ is calculated . The target group size is set to three loads. The resulting groups are shown in \cref{fig:clust_e_pmax1}. 

\subsection{Features choice for \smanet}

\begin{figure}[H]
    \centering
    \subfloat[\label{fig:e_clust} Energy]{\includegraphics[width=.5 \columnwidth]{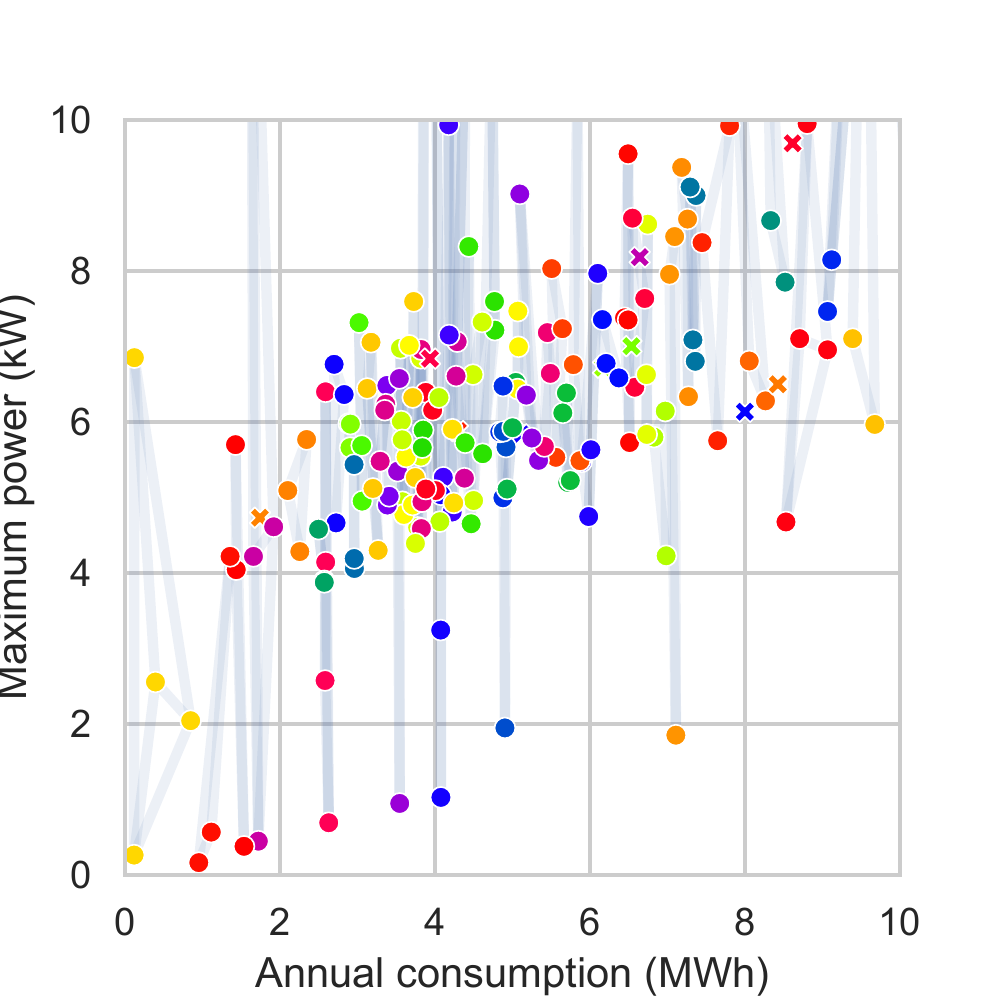}}
    \subfloat[\label{fig:pca_clust} PCA]{\includegraphics[width=.5 \columnwidth]{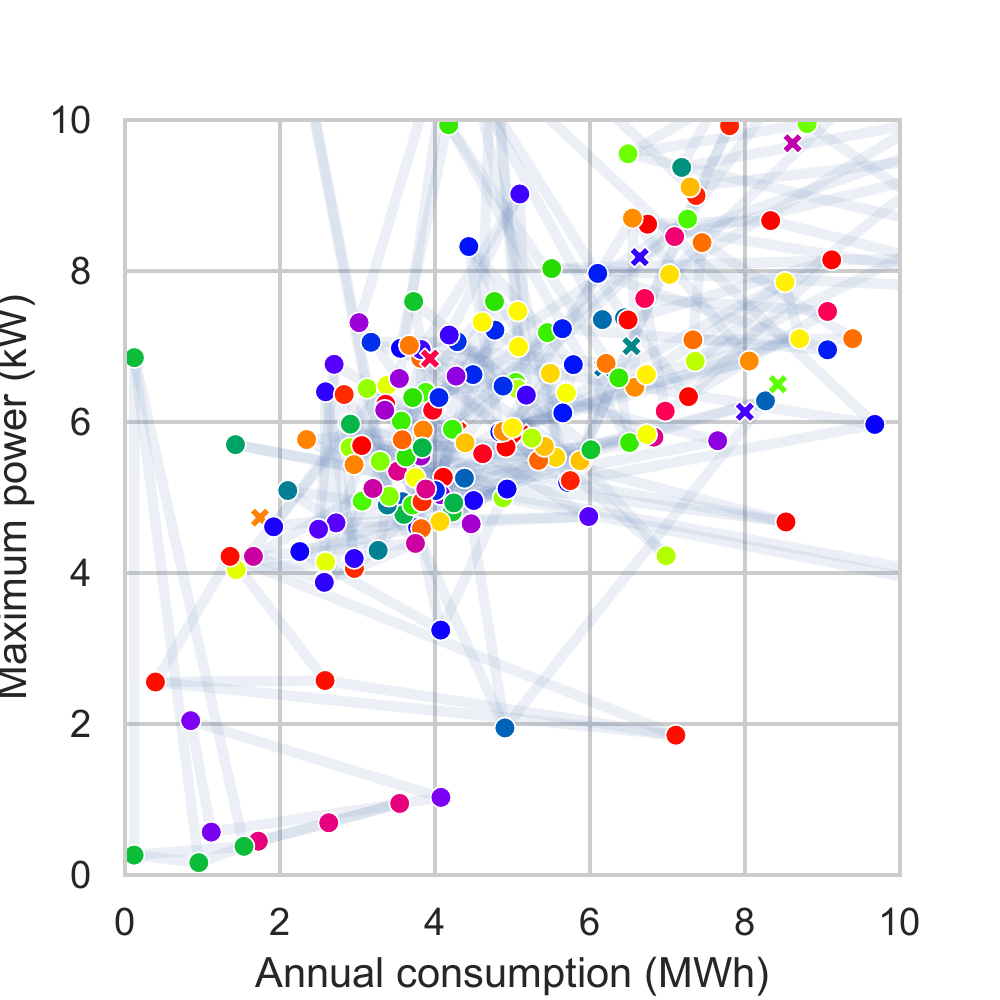}}\\
    \subfloat[\label{fig:dummy_clust} Affinity]{\includegraphics[width=.5 \columnwidth]{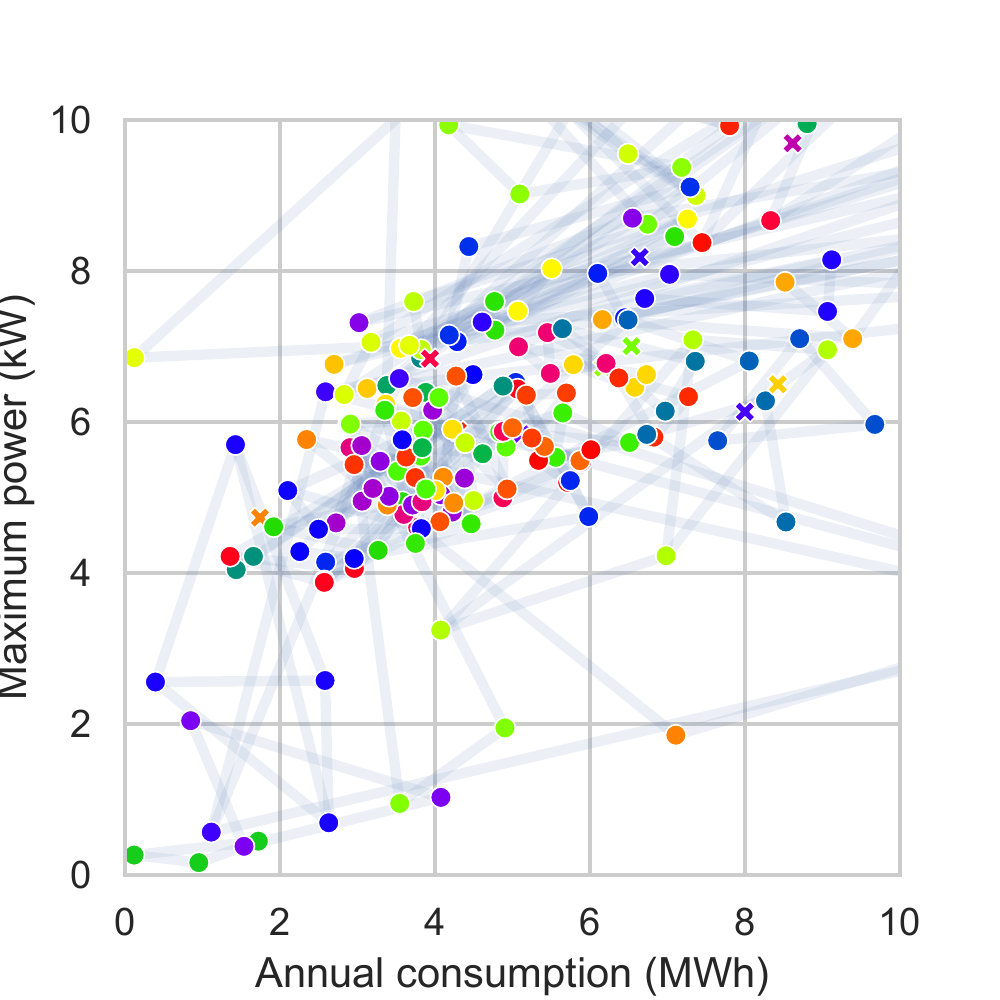}}
    \subfloat[\label{fig:one_group} One group]{\includegraphics[width=.5 \columnwidth]{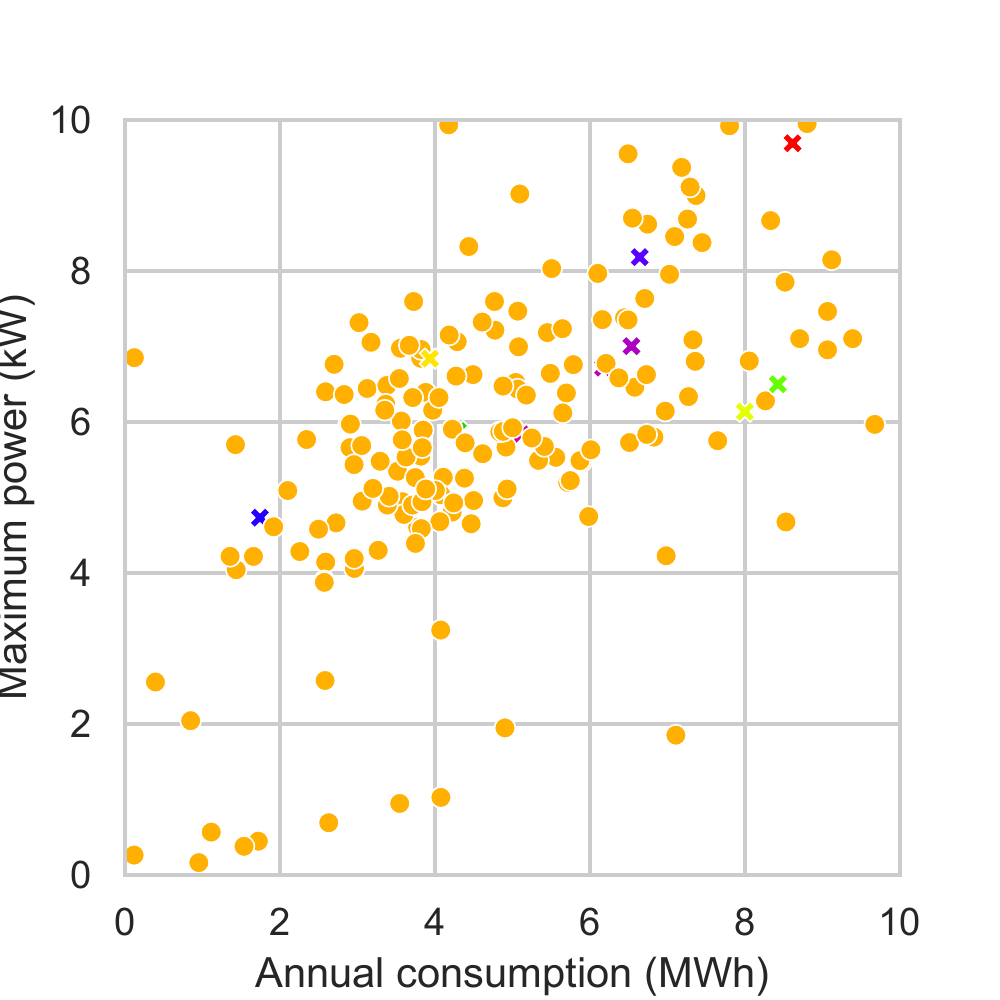}}\\
    \caption{Four additional partitioning scenarios. Measurements with more than three customers are marked with a $\times$.}
    \label{fig:clust_scen}
\end{figure}

In a second approach, we measure the impact of the feature choice on the load-flow solution's accuracy. To do so, we define five  partitioning scenarios:
\begin{description}
    \item[Energy and maximum power] These are the same features as defined in \cref{sec:sm_anony_method}, hereafter shortened ''E + max P''.
    \item[Energy] Only the annual energy consumption is considered.
    \item[PCA] From a set of features proposed by \cite{Beckel2012}, we perform a principal component analysis (PCA) and keep the first $N$ components that explain 99\% of the dataset variance.
    \item[Affinity] Again, E + max P are used as input features, but with the affinity matrix ($A_{i,j} = 1/D_{i,j}$) instead of the distance matrix. In the RSGP, this would be equivalent to finding the partitioning with the largest intra-cluster variance. 
    \item[One group] Instead of grouping the loads by three, all (except those measuring more than three customers) are put into a single group.
\end{description}
The resulting four additional partitioning scenarios are illustrated in \cref{fig:clust_scen}. In these figures, we kept the projection on the energy - max Power plane. This can lead to unnatural cluster representations as for the PCA scenarios (\cref{fig:pca_clust}). Note that for the Energy scenarios (\cref{fig:e_clust}), the clusters are formed by vertical slicing of the dataset.  

As described in the workflow of the \smanet methodology in \cref{fig:smanet_wf}, the final stage is to randomly select one permutation of the buses per group and allocate each load to its bus. In this stage, to account for the stochastic nature of this method, the load-flow problem is solved 200 times, each time with a new a new permutation of the bus-load assignment. 

%%%%%%%%%%%%%%%%%%%%%%%%%%%%%%%%%%%%%%%%%%%%%%%%%%%%%%%%%%%%%%
%%%%%%%%%%%%%%%%%%%%%%%%%%%%%%%%%%%%%%%%%%%%%%%%%%%%%%%%%%%%%%
%%%%%%%%%%%%%%%%%%%%%%%%%%%%%%%%%%%%%%%%%%%%%%%%%%%%%%%%%%%%%%
\section{Results}
\label{sec:results}

\subsection{Load allocation and \smanet comparison}

\begin{figure}[H]
    \centering
    \includegraphics[width=.7 \columnwidth]{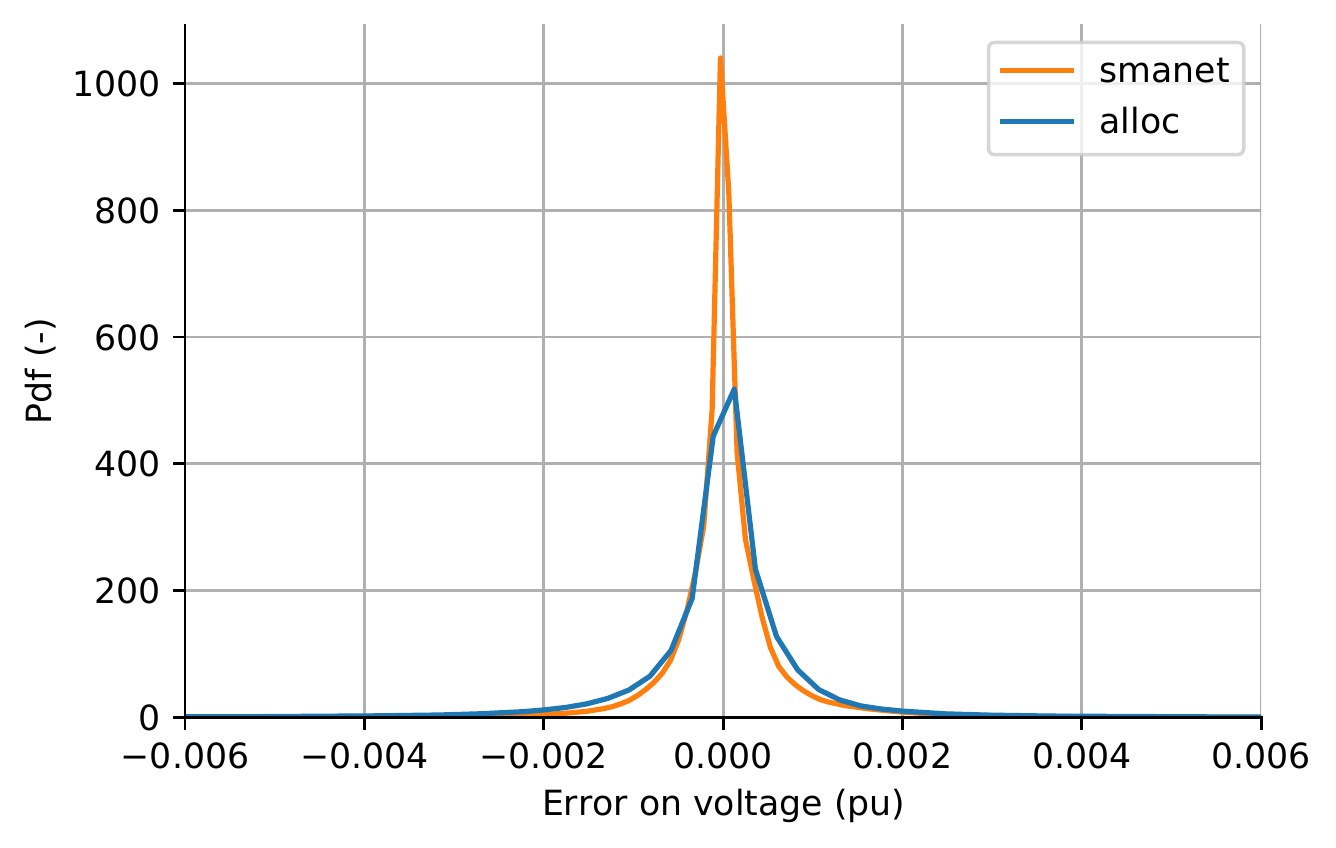}
    \caption{Voltage error distribution }
    \label{fig:V_error}
\end{figure}

The voltage error distribution across all times and all buses in the six networks are pictured in \cref{fig:V_error}. Here, only a single random permutation is used for the \smanet method. (The allocation provides by definition only one solution.) Both the allocation and the \smanet method give errors mostly below 0.002\,pu, which are already sufficient for most network studies. The \smanet process seems to provide slightly smaller errors.
These minor errors can be explained by the fact that during a large portion of the year, the active power demand is small compared to the network capacity (during the night, for instance), leading to a local voltage close to 1\,pu. 

\begin{figure}[H]
    \centering
    \includegraphics[width=.7 \columnwidth]{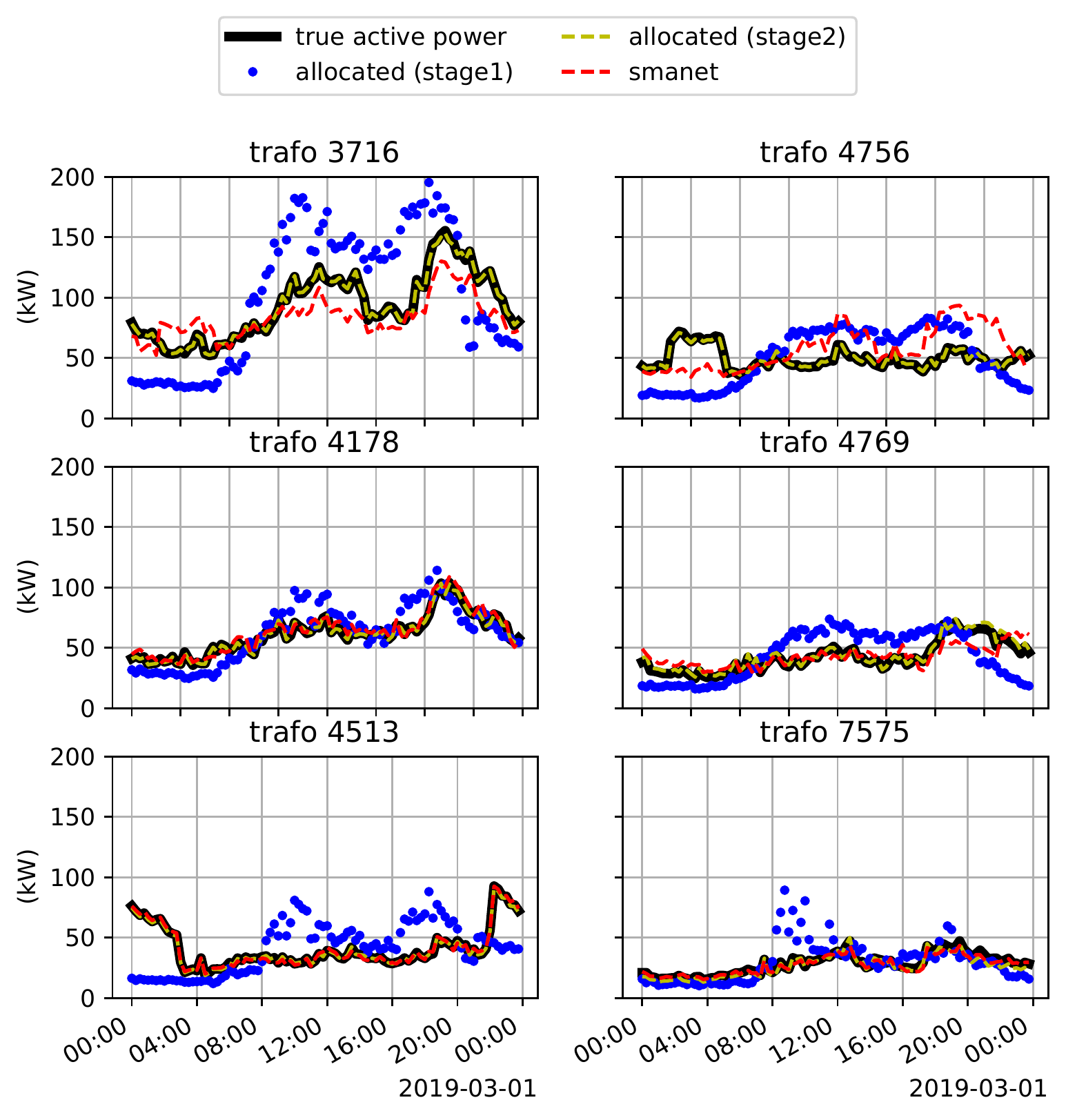}
    \caption{Active power at the transformer nodes of the six low-voltage grids}
    \label{fig:Ptrafo}
\end{figure}

To balance this effect, one must look at what is happening at the transformer nodes. The transformer's active power is plotted for the six networks for a particular day in \cref{fig:Ptrafo}. This figure shows how the second-stage optimization of the allocation method improves transformer state estimation accuracy compared with the first-stage obtained power (blue dots). By definition, the maximum power deviation at the transformer should be smaller than 1\%. For this reason, the allocated (stage 2) curve is very close to the true one. The resulting power at the transformer obtained with the \smanet method also leads to very good results. The quality of this method also lies in the fraction of the network loads given by the measure of more than the required three customers. In other words, for a network where all loads are multi-family buildings with more than three apartments, all loads' locations are known, and the network state estimation is very accurate (and there is no need for the anonymisation methodology). 

\begin{figure}[H]
\centering
    \subfloat[Fully allocated load\label{fig:P_bus1}]{\includegraphics[width=.7\columnwidth]{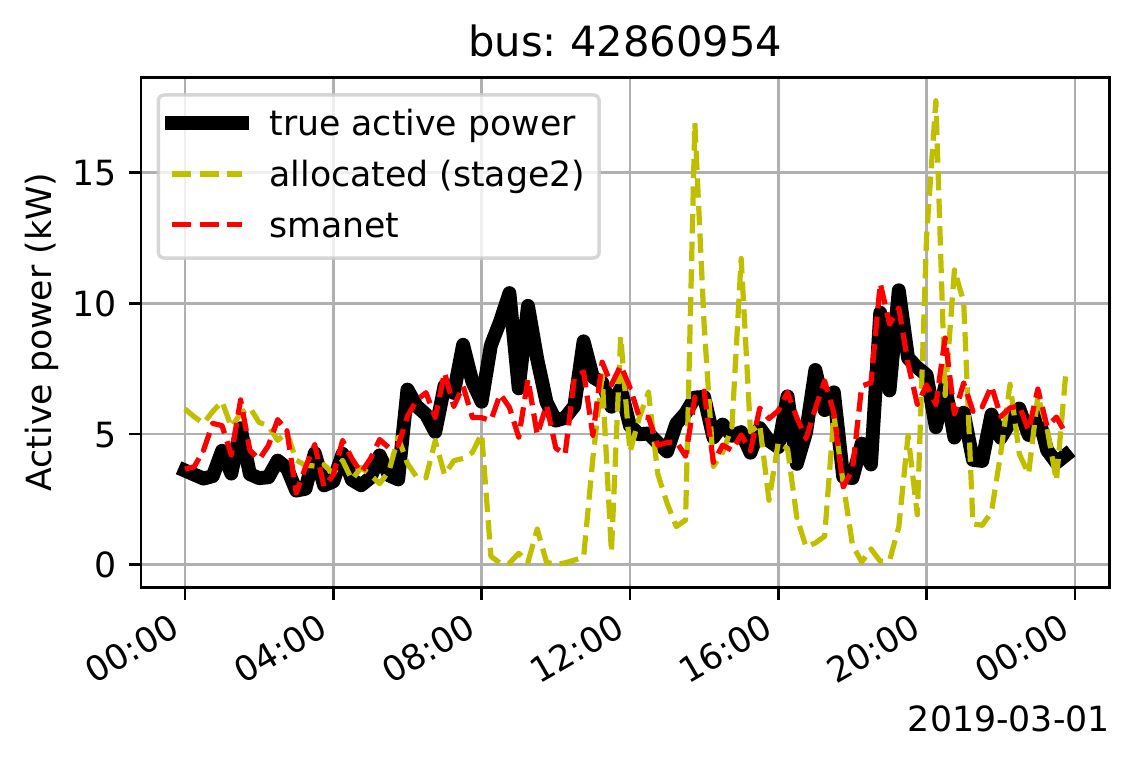}}
    \\
    \subfloat[\label{fig:P_bus2}Naturally anonymised load]{\includegraphics[width=.7\columnwidth]{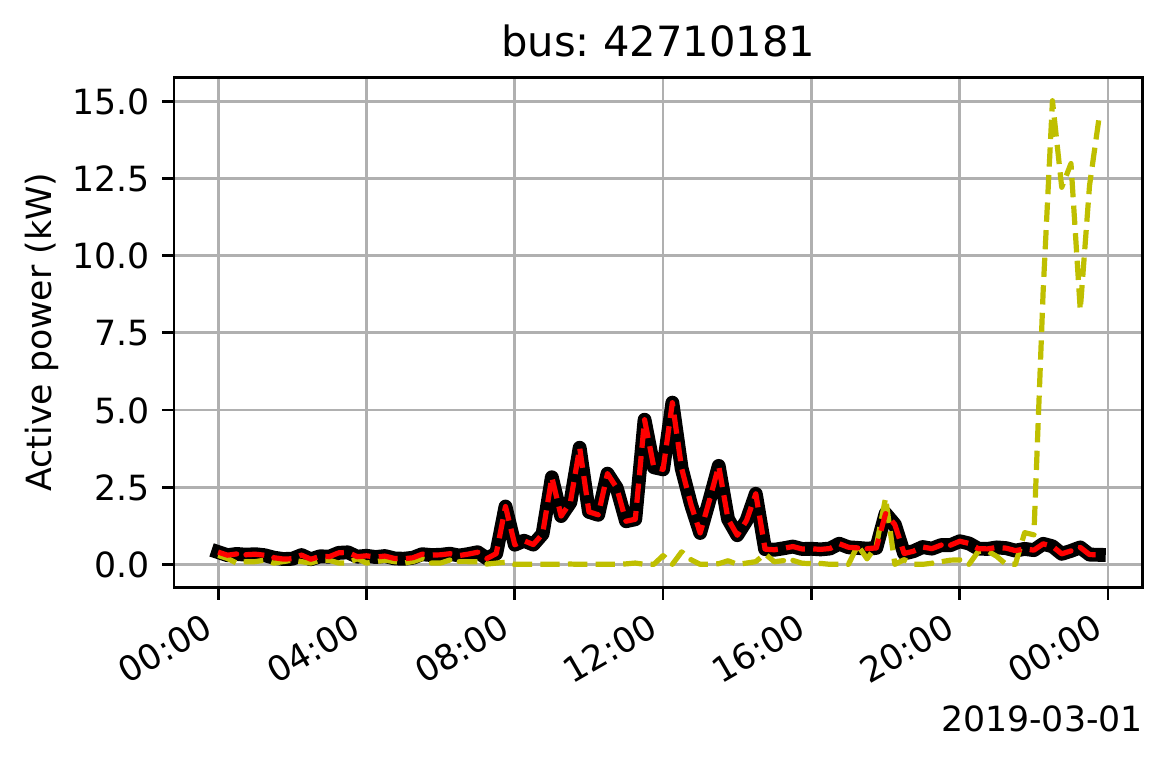}}
    \caption{Example of active power at two buses }
    \label{fig:P_bus}
\end{figure}

Despite the good voltage accuracy for both methods, the local power allocated to a given network location (bus) is closer to the true one with the \smanet method than with the allocation method (\cref{fig:P_bus}). The illustrative example of \cref{fig:P_bus2} shows that the \smanet benefits from the network location knowledge in some cases. It also happens that by chance (1 out of 3), the load allocated to this bus is the actual original load of this bus.

Finally, the key performance indicators are reported in \cref{tab:alloc_kpi}. Again thanks to prior knowledge of the load locations, the \smanet method is slightly more accurate. The minimum voltage estimation error is below 3\% for both methods. The maximum transformer loading is underestimated by 19\% for the \smanet methods versus 1.2\% for the allocation methods. The advantage of the allocation method on this metric is the constraints on the transformer's power that should be below 1\% in this case. The additional 0.2\% comes from the fact that no prior knowledge of the grid losses is used in the allocation (the transformer's power is assumed to be the sum of the network loads, neglecting the line losses).
The accuracy of the \smanet methods also depends on the loads' final allocation, i.e. for a given bus, out of the three loads belonging to the corresponding group, which load is attributed to the bus. In addition, the loads' intra-group similarity is critical to have accurate network state estimation. In the following, the latter is discussed by evaluating the five partitioning scenarios and running, for each scenario, the load-flow simulation 200 times. 

\newcommand{\tento}[2]{\ensuremath{{{#1} \cdot 10^{#2}}}}
% \begin{table}[H]
%     \centering
%      \caption{Performance indicators}
%     \label{tab:alloc_kpi}
%     \begin{tabular}{@{}lcc@{}}
%     \toprule
%  & \textbf{allocation} & \textbf{\smanet} \\
%  \midrule
% $\tf{MSE}_\tf{vm}$      & \tento{6.33}{-7}  & \tento{3.20}{-7}\\
% $\tf{E}_\tf{maxTRL}$    & 1.23 \%           & -18.90 \% \\
% $\tf{E}_\tf{maxLNL} $   & 1.83              & \tento{-5.46}{-5} \\
% $\tf{E}_\tf{minVM}$     & -2.76 \%          & 0.16 \% \\
% \bottomrule
%     \end{tabular}
   
% \end{table}

\begin{table}[H]
\centering
     \caption{Performance indicators per network for the allocation (A) and \smanet (S) methods}
    \label{tab:alloc_kpi}
\begin{tabular}{@{}ll|cccccc@{}}
\toprule
\multicolumn{2}{r}{\textbf{Network ID:}}                          &  \textbf{3716} &  \textbf{4178} &  \textbf{4513} &  \textbf{4756} &  \textbf{4769} &  \textbf{7575} \\
                         \midrule
\multirow{2}{*}{$\tf{MSE}_\tf{vm}\,(\cdot 10^{-7}\,\tf{pu}^2)$} & A   &
16.7 &1.3 &24.7 & 2.7 & 2.3 &	1.6\\
                                    & S      &
6.8	& 0.5& 	0.6 &	8.0 &	2.0 &	0.3\\ \midrule
\multirow{2}{*}{$\tf{E}_\tf{maxTRL}$ (\%) } & A &
1.2  &	-0.9  &	-0.5  &	-0.2  &	0.1  &	0.0  \\
                                         & S    &
-18.9  &	10.4  &	1.1  &	32.3  &	6.8  &	2.9  \\ \midrule
\multirow{2}{*}{$\tf{E}_\tf{maxLNL} $ (\%) }  & A  &
192.0  &	304.2  &	52.9  &	207.6  &	212.4  &	91.6  \\
                                    &    S  &
2.4  &	0.0  &	0.0  &	-3.0  &	15.6  &	8.4  \\ \midrule
\multirow{2}{*}{$\tf{E}_\tf{minVM}\,(\cdot 10^{-3}\,\tf{pu})$ } & A   &
-27.0 &	-13.9 &	-5.6 &	-13.7 &	-4.9 &	-3.6 \\
                                    & S         &
1.6 &	-0.7 &	-0.6 &	-0.9 &	-4.2 &	-0.1\\ \bottomrule
\end{tabular}
\end{table}

\subsection{Features' influence on \smanet accuracy}
\begin{figure}[H]
    \centering
    \includegraphics[width=.8 \columnwidth]{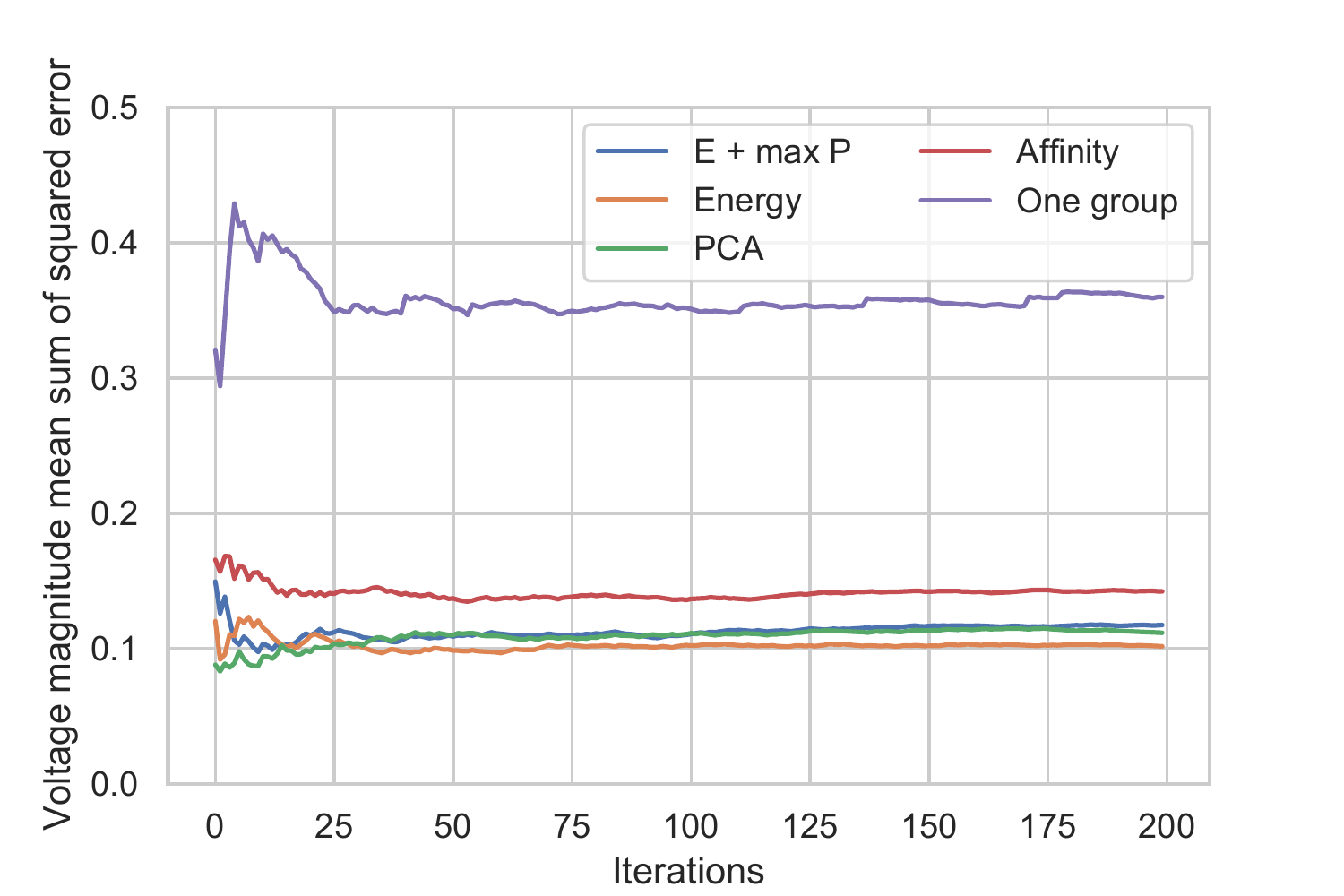}
    \caption{Voltage magnitude mean squared error convergence}
    \label{fig:MSEconv}
\end{figure}

As mentioned, to have a clear overview of the \smanet accuracy, the load flows are solved several times with a new load-bus assignment. The key performance indicators are recorded for each iteration. The voltage magnitude mean squared error (calculated for all buses at all times for all iterations until the $i^{th}$ iteration) is plotted in \cref{fig:MSEconv}. This figure shows the convergence of the mean squared error for all partitioning scenarios.  

\begin{figure}[H]
    \centering
    \includegraphics[width=.8 \columnwidth]{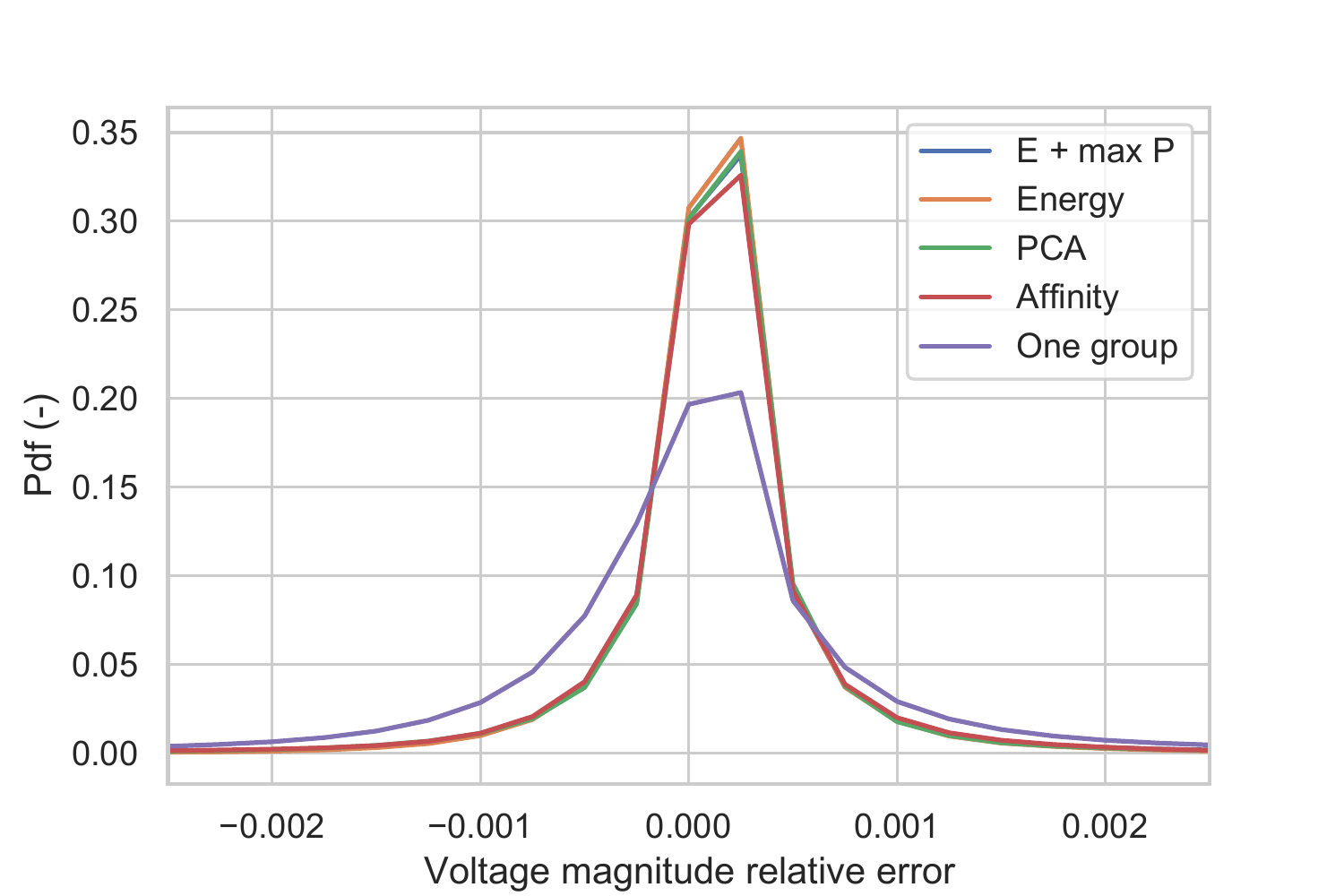}
    \caption{Error on voltage magnitude }
    \label{fig:vm_error_RE}
\end{figure}

The voltage magnitude error (across all times, buses, and iterations) is plotted in \cref{fig:vm_error_RE}. Again the error is mostly smaller than 0.002\,pu. Except for the Dummy and One group scenarios, the three other scenarios provide similar accuracy from this perspective. 

The mean squared error of the voltage magnitude and its standard deviation (across iterations) are reported in \cref{tab:MSE_smanet}. Again no significant differences are observed between scenarios, except for the Affinity and One group scenarios that give slightly larger mean squared errors of the voltage magnitude.

\begin{table}[H]
    \centering
    \caption{Mean squared error of the voltage magnitude for the five partitioning scenarios}
    \label{tab:MSE_smanet}
    \begin{tabular}{@{}lccccc@{}}
    \toprule
    {} & \textbf{E + max P} &    \textbf{Energy} &       \textbf{PCA} & \textbf{Affinity} &    \textbf{One group} \\
    \midrule
    \textbf{Mean} &    \tento{2.02}{-8} &  \tento{1.70}{-8} &  \tento{2.02}{-8} &\tento{2.24}{-8} &  \tento{3.92}{-8} \\
    \textbf{Std}  &   \tento{7.66}{-9} &  \tento{6.84}{-9} &  \tento{9.05}{-9} &  \tento{7.18}{-9} &  \tento{1.68}{-8} \\
    \bottomrule
    \end{tabular}
\end{table}

%%%%%%%%%%%%%%%%%%%%%%%%%%%%%%%%%%%%%%%%%%%%%%%%%%%%%%%%%%%%%%%%%%%%%%%%%
%%%%%%%%%%%%%%%%%%%%%%%%%%%%%%%%%%%%%%%%%%%%%%%%%%%%%%%%%%%%%%%%%%%%%%%%%
%%%%%%%%%%%%%%%%%%%%%%%%%%%%%%%%%%%%%%%%%%%%%%%%%%%%%%%%%%%%%%%%%%%%%%%%%
\section{Discussion}
\label{sec:discuss}

The results highlight that both methods provide similar results in terms of network state estimation accuracy. In more detail, the \smanet approach provides slightly higher accuracy with respect to voltage estimation, which translates into higher accuracy in terms of line loading. The assumption that a load can be exchanged by an electrically similar load while still keeping accurate network simulations seems valid in this case. The allocation method provides better results in terms of transformer loading. The reason behind this observation is the applied constraints on the estimated active power at the transformer side. This linear constraint does not account for line losses nor for line inductance, which requires the provision of reactive power. Still, the results show that this does not have a significant impact.  The \smanet approach suffers from a significant error in transformer loading. This is in contrast with its low estimation error on voltage and line loading.   The \smanet approach should be preferred as soon as SM data are available inside a given network. When this is not the case, the allocation of SM measurements from an external network is the only suitable solution. The allocation and \smanet approaches could be combined. For the network locations where SMs are installed, the \smanet approach could be used. For the network locations where no SMs are currently set up, SM data acquired outside the considered networks could be allocated to these locations. In such a case, it would be better to use load profiles from another network than to just reuse the one measured in the same network to avoid simultaneous peaks. However, if not possible, the second-stage optimization of the allocation method should reduce this issue by deforming the loads. 

Both approaches rely on the fact that SMs are considered anonymous if the true location of the original meter is not known. However, the allocation approach relies on labeling to define load categories. The numbers of these categories could be extended, but it might raise issues if the label gets so precise that it allows the true origin of the meter to be linked with the SM measurements. These categories could be based on publicly available data regarding buildings, e.g. from the cantonal/federal building registries. The registries contain more information on the buildings that could improve the classification of the loads, thus the accuracy of the network simulation.  The degree of anonymisation in the \smanet approach becomes questionable when groups with a large intra-distance occur, i.e. if a group contains three very significantly different annual energy consumption values. In this case, given the fact that three possible locations for these loads are provided, it might be easy to deduce the true location of the loads, as pointed out by \cite{Eibl2021}. This can be addressed first by increasing the dataset size, second by increasing the group size, and finally by excluding those loads from the dataset. Besides, in this work, we consider only six low-voltage networks and the corresponding installed SMs. The extension of the \smanet methodology to an entire DSO control area raises the following question: Should all SM data be encompassed in a single dataset, or is it possible to split the dataset by geographical area? 
In this work, we consider only one sufficiently large dataset and discussed the limitation of such methodologies. It might be worth combining the knowledge gained from information theory to tackle this question.

%%%%%%%%%%%%%%%%%%%%%%%%%%%%%%%%%%%%%%%%%%%%%%%%%%%%%%%%%%%%%%%%%%%%%%%%%
%%%%%%%%%%%%%%%%%%%%%%%%%%%%%%%%%%%%%%%%%%%%%%%%%%%%%%%%%%%%%%%%%%%%%%%%%
%%%%%%%%%%%%%%%%%%%%%%%%%%%%%%%%%%%%%%%%%%%%%%%%%%%%%%%%%%%%%%%%%%%%%%%%%
\section{Conclusion}
\label{sec:conclusion}

In this work, two methods to use SMs for network simulations were proposed. The first method was based on an allocation approach. First, the allocation was formulated as a mixed-integer problem to allocate loads from a database to the network locations based on the annual energy demand difference. Second, an adjustment stage deformed the original loads to match the power at the transformer. 

The second method aimed to anonymise the smart measurements extracted from a given set of networks by grouping the SM loads by $K$  and linking this group to a set of $K$ network locations (buses). The final load-bus allocation was achieved by randomly assigning a load to a bus in the considered group. This method allowed for stochastic analysis of the network stage (by randomly permuting the assignment). A stochastic approach improved the quality of the network simulation but at a higher computational burden. This approach showed the best accuracy with respect to state estimation and voltage magnitude estimation. Line loading errors were significant as they result from an accumulation of downstream load errors.

Both methods allow for an anonymised usage of SM data for network simulations. Thus they are particularly useful for DSOs. Combining both methods could prove useful. As the rollout of SMs is just starting, some networks might contain very few or no SMs at all. A combination of both approaches could allow for a smooth transition from a low SM density to a future with all customers equipped with SMs. 

%%%%%%%%%%%%%%%%%%%%%%%%%%%%%%%%%%%%%%%%%%%%%%%%%%%%%%%%%%%%%%%%%%%%%%%%%
%%%%%%%%%%%%%%%%%%%%%%%%%%%%%%%%%%%%%%%%%%%%%%%%%%%%%%%%%%%%%%%%%%%%%%%%%
%%%%%%%%%%%%%%%%%%%%%%%%%%%%%%%%%%%%%%%%%%%%%%%%%%%%%%%%%%%%%%%%%%%%%%%%%
\section{Acknowledgement}
This work was supported by InnoSuisse in the
framework of the SCCER-FURIES. We would like to thank the team from \textit{Romande Energie}, Patrizio Canzi,  Kim Leng Chhun, Assia Garbinato, Tiago Torrado, and Arnoud Bifrare for their help and precious support.

%\appendix

%\section{Sample Appendix Section}
%\lipsum

 \bibliographystyle{elsarticle-num} 
 \bibliography{biblio}





% \end{thebibliography}
\end{document}

%% file: nomenclature.tex
\usepackage{xspace}

\newcommand{\pv}{\text{PV}\xspace}
\newcommand{\flexi}{\textsc{flexi}\xspace}
\newcommand{\flexii}{\textsc{flexi 2}\xspace}
\newcommand{\smanet}{\textsc{smanet}\xspace}
\newcommand{\RE}{\textit{Romande Energy}\xspace}

\newcommand{\tf}{\mathrm} % text format

\newcommand{\ts}{\ensuremath{TS}}
\newcommand{\Reff}{\ensuremath{\tf{ref}}}
\newcommand{\Org}{\ensuremath{\tf{org}}}
\newcommand{\Var}{\ensuremath{\tf{var}}}

\newcommand{\pvar}{\ensuremath{P^\Var}}
\newcommand{\pref}{\ensuremath{P^\Reff}}
\newcommand{\porg}{\ensuremath{P^\Org}}
\newcommand{\evar}{\ensuremath{E^\Var}}
\newcommand{\eref}{\ensuremath{E^\Reff}}
\newcommand{\eorg}{\ensuremath{E^\Org}}

%% file: figure/NetworkIllust.tex
% Graphic for TeX using PGF
% Title: C:\Users\holweger\switchdrive\Doctorat PVlab\Projets\disagregation\load-allocations\NetworkIllust.dia
% Creator: Dia v0.97.2
% CreationDate: Fri Jan 11 10:12:02 2019
% For: holweger
% \usepackage{tikz}
% The following commands are not supported in PSTricks at present
% We define them conditionally, so when they are implemented,
% this pgf file will use them.
\ifx\du\undefined
  \newlength{\du}
\fi
\setlength{\du}{15\unitlength}
\begin{tikzpicture}
\pgftransformxscale{1.000000}
\pgftransformyscale{-1.000000}
\definecolor{dialinecolor}{rgb}{0.000000, 0.000000, 0.000000}
\pgfsetstrokecolor{dialinecolor}
\definecolor{dialinecolor}{rgb}{1.000000, 1.000000, 1.000000}
\pgfsetfillcolor{dialinecolor}
\definecolor{dialinecolor}{rgb}{0.000000, 1.000000, 0.000000}
\pgfsetfillcolor{dialinecolor}
\pgfpathellipse{\pgfpoint{12.727456\du}{2.150000\du}}{\pgfpoint{1.000000\du}{0\du}}{\pgfpoint{0\du}{1.000000\du}}
\pgfusepath{fill}
\pgfsetlinewidth{0.100000\du}
\pgfsetdash{}{0pt}
\pgfsetdash{}{0pt}
\definecolor{dialinecolor}{rgb}{0.000000, 0.000000, 0.000000}
\pgfsetstrokecolor{dialinecolor}
\pgfpathellipse{\pgfpoint{12.727456\du}{2.150000\du}}{\pgfpoint{1.000000\du}{0\du}}{\pgfpoint{0\du}{1.000000\du}}
\pgfusepath{stroke}
\pgfsetlinewidth{0.100000\du}
\pgfsetdash{}{0pt}
\pgfsetdash{}{0pt}
\pgfsetbuttcap
{
\definecolor{dialinecolor}{rgb}{0.000000, 0.000000, 0.000000}
\pgfsetfillcolor{dialinecolor}
% was here!!!
\pgfsetarrowsend{to}
\definecolor{dialinecolor}{rgb}{0.000000, 0.000000, 0.000000}
\pgfsetstrokecolor{dialinecolor}
\draw (12.727456\du,3.150000\du)--(1.997500\du,6.544911\du);
}
\pgfsetlinewidth{0.100000\du}
\pgfsetdash{}{0pt}
\pgfsetdash{}{0pt}
\pgfsetbuttcap
{
\definecolor{dialinecolor}{rgb}{0.000000, 0.000000, 0.000000}
\pgfsetfillcolor{dialinecolor}
% was here!!!
\pgfsetarrowsend{to}
\definecolor{dialinecolor}{rgb}{0.000000, 0.000000, 0.000000}
\pgfsetstrokecolor{dialinecolor}
\draw (12.727456\du,3.150000\du)--(5.574152\du,6.544911\du);
}
\pgfsetlinewidth{0.100000\du}
\pgfsetdash{}{0pt}
\pgfsetdash{}{0pt}
\pgfsetbuttcap
{
\definecolor{dialinecolor}{rgb}{0.000000, 0.000000, 0.000000}
\pgfsetfillcolor{dialinecolor}
% was here!!!
\pgfsetarrowsend{to}
\definecolor{dialinecolor}{rgb}{0.000000, 0.000000, 0.000000}
\pgfsetstrokecolor{dialinecolor}
\draw (12.727456\du,3.150000\du)--(9.150804\du,6.544911\du);
}
\pgfsetlinewidth{0.100000\du}
\pgfsetdash{}{0pt}
\pgfsetdash{}{0pt}
\pgfsetbuttcap
{
\definecolor{dialinecolor}{rgb}{0.000000, 0.000000, 0.000000}
\pgfsetfillcolor{dialinecolor}
% was here!!!
\pgfsetarrowsend{to}
\definecolor{dialinecolor}{rgb}{0.000000, 0.000000, 0.000000}
\pgfsetstrokecolor{dialinecolor}
\draw (12.727456\du,3.150000\du)--(12.727456\du,6.544911\du);
}
\pgfsetlinewidth{0.100000\du}
\pgfsetdash{}{0pt}
\pgfsetdash{}{0pt}
\pgfsetbuttcap
{
\definecolor{dialinecolor}{rgb}{0.000000, 0.000000, 0.000000}
\pgfsetfillcolor{dialinecolor}
% was here!!!
\pgfsetarrowsend{to}
\definecolor{dialinecolor}{rgb}{0.000000, 0.000000, 0.000000}
\pgfsetstrokecolor{dialinecolor}
\draw (12.727456\du,3.150000\du)--(16.304108\du,6.544911\du);
}
\pgfsetlinewidth{0.100000\du}
\pgfsetdash{}{0pt}
\pgfsetdash{}{0pt}
\pgfsetbuttcap
{
\definecolor{dialinecolor}{rgb}{0.000000, 0.000000, 0.000000}
\pgfsetfillcolor{dialinecolor}
% was here!!!
\pgfsetarrowsend{to}
\definecolor{dialinecolor}{rgb}{0.000000, 0.000000, 0.000000}
\pgfsetstrokecolor{dialinecolor}
\draw (12.727456\du,3.150000\du)--(19.880759\du,6.544911\du);
}
\pgfsetlinewidth{0.100000\du}
\pgfsetdash{}{0pt}
\pgfsetdash{}{0pt}
\pgfsetbuttcap
{
\definecolor{dialinecolor}{rgb}{0.000000, 0.000000, 0.000000}
\pgfsetfillcolor{dialinecolor}
% was here!!!
\pgfsetarrowsend{to}
\definecolor{dialinecolor}{rgb}{0.000000, 0.000000, 0.000000}
\pgfsetstrokecolor{dialinecolor}
\draw (12.727456\du,3.150000\du)--(23.457411\du,6.544911\du);
}
\pgfsetlinewidth{0.100000\du}
\pgfsetdash{{1.000000\du}{1.000000\du}}{0\du}
\pgfsetdash{{1.000000\du}{1.000000\du}}{0\du}
\pgfsetbuttcap
{
\definecolor{dialinecolor}{rgb}{0.000000, 0.000000, 0.000000}
\pgfsetfillcolor{dialinecolor}
% was here!!!
\pgfsetarrowsend{to}
\definecolor{dialinecolor}{rgb}{0.000000, 0.000000, 0.000000}
\pgfsetstrokecolor{dialinecolor}
\draw (1.997500\du,8.544911\du)--(12.727456\du,13.300000\du);
}
\pgfsetlinewidth{0.100000\du}
\pgfsetdash{{1.000000\du}{1.000000\du}}{0\du}
\pgfsetdash{{1.000000\du}{1.000000\du}}{0\du}
\pgfsetbuttcap
{
\definecolor{dialinecolor}{rgb}{0.000000, 0.000000, 0.000000}
\pgfsetfillcolor{dialinecolor}
% was here!!!
\pgfsetarrowsend{to}
\definecolor{dialinecolor}{rgb}{0.000000, 0.000000, 0.000000}
\pgfsetstrokecolor{dialinecolor}
\draw (5.574152\du,8.544911\du)--(1.997500\du,13.300000\du);
}
\pgfsetlinewidth{0.100000\du}
\pgfsetdash{{1.000000\du}{1.000000\du}}{0\du}
\pgfsetdash{{1.000000\du}{1.000000\du}}{0\du}
\pgfsetbuttcap
{
\definecolor{dialinecolor}{rgb}{0.000000, 0.000000, 0.000000}
\pgfsetfillcolor{dialinecolor}
% was here!!!
\pgfsetarrowsend{to}
\definecolor{dialinecolor}{rgb}{0.000000, 0.000000, 0.000000}
\pgfsetstrokecolor{dialinecolor}
\draw (23.457411\du,8.544911\du)--(18.092433\du,13.300000\du);
}
\pgfsetlinewidth{0.100000\du}
\pgfsetdash{{1.000000\du}{1.000000\du}}{0\du}
\pgfsetdash{{1.000000\du}{1.000000\du}}{0\du}
\pgfsetbuttcap
{
\definecolor{dialinecolor}{rgb}{0.000000, 0.000000, 0.000000}
\pgfsetfillcolor{dialinecolor}
% was here!!!
\pgfsetarrowsend{to}
\definecolor{dialinecolor}{rgb}{0.000000, 0.000000, 0.000000}
\pgfsetstrokecolor{dialinecolor}
\draw (16.304108\du,8.544911\du)--(12.727456\du,13.300000\du);
}
\pgfsetlinewidth{0.100000\du}
\pgfsetdash{{1.000000\du}{1.000000\du}}{0\du}
\pgfsetdash{{1.000000\du}{1.000000\du}}{0\du}
\pgfsetbuttcap
{
\definecolor{dialinecolor}{rgb}{0.000000, 0.000000, 0.000000}
\pgfsetfillcolor{dialinecolor}
% was here!!!
\pgfsetarrowsend{to}
\definecolor{dialinecolor}{rgb}{0.000000, 0.000000, 0.000000}
\pgfsetstrokecolor{dialinecolor}
\draw (9.150804\du,8.544911\du)--(7.362478\du,13.300000\du);
}
\pgfsetlinewidth{0.100000\du}
\pgfsetdash{{1.000000\du}{1.000000\du}}{0\du}
\pgfsetdash{{1.000000\du}{1.000000\du}}{0\du}
\pgfsetbuttcap
{
\definecolor{dialinecolor}{rgb}{0.000000, 0.000000, 0.000000}
\pgfsetfillcolor{dialinecolor}
% was here!!!
\pgfsetarrowsend{to}
\definecolor{dialinecolor}{rgb}{0.000000, 0.000000, 0.000000}
\pgfsetstrokecolor{dialinecolor}
\draw (19.880759\du,8.544911\du)--(23.457411\du,13.300000\du);
}
% setfont left to latex
\definecolor{dialinecolor}{rgb}{0.000000, 0.000000, 0.000000}
\pgfsetstrokecolor{dialinecolor}
\node[anchor=west] at (10.802456\du,21.650000\du){$N_P$};
% setfont left to latex
\definecolor{dialinecolor}{rgb}{0.000000, 0.000000, 0.000000}
\pgfsetstrokecolor{dialinecolor}
\node[anchor=west] at (9.802456\du,21.650000\du){};
% setfont left to latex
\definecolor{dialinecolor}{rgb}{0.000000, 0.000000, 0.000000}
\pgfsetstrokecolor{dialinecolor}
\node[anchor=west] at (10.802456\du,18.350000\du){$N_L$};
% setfont left to latex
\definecolor{dialinecolor}{rgb}{0.000000, 0.000000, 0.000000}
\pgfsetstrokecolor{dialinecolor}
\node[anchor=west] at (4.400000\du,16.850000\du){};
% setfont left to latex
\definecolor{dialinecolor}{rgb}{0.000000, 0.000000, 0.000000}
\pgfsetstrokecolor{dialinecolor}
\node[anchor=west] at (9.950000\du,17.150000\du){};
% setfont left to latex
\definecolor{dialinecolor}{rgb}{0.000000, 0.000000, 0.000000}
\pgfsetstrokecolor{dialinecolor}
\node[anchor=west] at (16.652456\du,18.350000\du){$N_K$};
% setfont left to latex
\definecolor{dialinecolor}{rgb}{0.000000, 0.000000, 0.000000}
\pgfsetstrokecolor{dialinecolor}
\node[anchor=west] at (16.000000\du,20.350000\du){};
% setfont left to latex
\definecolor{dialinecolor}{rgb}{0.000000, 0.000000, 0.000000}
\pgfsetstrokecolor{dialinecolor}
\node[anchor=west] at (16.652456\du,21.650000\du){$J$};
% setfont left to latex
\definecolor{dialinecolor}{rgb}{0.000000, 0.000000, 0.000000}
\pgfsetstrokecolor{dialinecolor}
\node[anchor=west] at (18.500000\du,16.650000\du){};
% setfont left to latex
\definecolor{dialinecolor}{rgb}{0.000000, 0.000000, 0.000000}
\pgfsetstrokecolor{dialinecolor}
\node[anchor=west] at (4.700000\du,17.150000\du){};
\definecolor{dialinecolor}{rgb}{0.000000, 1.000000, 0.000000}
\pgfsetfillcolor{dialinecolor}
\pgfpathellipse{\pgfpoint{9.802456\du}{21.650000\du}}{\pgfpoint{1.000000\du}{0\du}}{\pgfpoint{0\du}{1.000000\du}}
\pgfusepath{fill}
\pgfsetlinewidth{0.100000\du}
\pgfsetdash{}{0pt}
\pgfsetdash{}{0pt}
\definecolor{dialinecolor}{rgb}{0.000000, 0.000000, 0.000000}
\pgfsetstrokecolor{dialinecolor}
\pgfpathellipse{\pgfpoint{9.802456\du}{21.650000\du}}{\pgfpoint{1.000000\du}{0\du}}{\pgfpoint{0\du}{1.000000\du}}
\pgfusepath{stroke}
\definecolor{dialinecolor}{rgb}{1.000000, 0.647059, 0.000000}
\pgfsetfillcolor{dialinecolor}
\pgfpathellipse{\pgfpoint{9.802456\du}{18.350000\du}}{\pgfpoint{1.000000\du}{0\du}}{\pgfpoint{0\du}{1.000000\du}}
\pgfusepath{fill}
\pgfsetlinewidth{0.100000\du}
\pgfsetdash{}{0pt}
\pgfsetdash{}{0pt}
\definecolor{dialinecolor}{rgb}{0.000000, 0.000000, 0.000000}
\pgfsetstrokecolor{dialinecolor}
\pgfpathellipse{\pgfpoint{9.802456\du}{18.350000\du}}{\pgfpoint{1.000000\du}{0\du}}{\pgfpoint{0\du}{1.000000\du}}
\pgfusepath{stroke}
\definecolor{dialinecolor}{rgb}{0.000000, 0.000000, 1.000000}
\pgfsetfillcolor{dialinecolor}
\pgfpathellipse{\pgfpoint{15.652456\du}{21.650000\du}}{\pgfpoint{1.000000\du}{0\du}}{\pgfpoint{0\du}{1.000000\du}}
\pgfusepath{fill}
\pgfsetlinewidth{0.100000\du}
\pgfsetdash{}{0pt}
\pgfsetdash{}{0pt}
\definecolor{dialinecolor}{rgb}{0.000000, 0.000000, 0.000000}
\pgfsetstrokecolor{dialinecolor}
\pgfpathellipse{\pgfpoint{15.652456\du}{21.650000\du}}{\pgfpoint{1.000000\du}{0\du}}{\pgfpoint{0\du}{1.000000\du}}
\pgfusepath{stroke}
\definecolor{dialinecolor}{rgb}{1.000000, 1.000000, 0.000000}
\pgfsetfillcolor{dialinecolor}
\pgfpathellipse{\pgfpoint{15.652456\du}{18.350000\du}}{\pgfpoint{1.000000\du}{0\du}}{\pgfpoint{0\du}{1.000000\du}}
\pgfusepath{fill}
\pgfsetlinewidth{0.100000\du}
\pgfsetdash{}{0pt}
\pgfsetdash{}{0pt}
\definecolor{dialinecolor}{rgb}{0.000000, 0.000000, 0.000000}
\pgfsetstrokecolor{dialinecolor}
\pgfpathellipse{\pgfpoint{15.652456\du}{18.350000\du}}{\pgfpoint{1.000000\du}{0\du}}{\pgfpoint{0\du}{1.000000\du}}
\pgfusepath{stroke}
\definecolor{dialinecolor}{rgb}{1.000000, 0.647059, 0.000000}
\pgfsetfillcolor{dialinecolor}
\pgfpathellipse{\pgfpoint{1.997500\du}{7.544911\du}}{\pgfpoint{1.000000\du}{0\du}}{\pgfpoint{0\du}{1.000000\du}}
\pgfusepath{fill}
\pgfsetlinewidth{0.100000\du}
\pgfsetdash{}{0pt}
\pgfsetdash{}{0pt}
\definecolor{dialinecolor}{rgb}{0.000000, 0.000000, 0.000000}
\pgfsetstrokecolor{dialinecolor}
\pgfpathellipse{\pgfpoint{1.997500\du}{7.544911\du}}{\pgfpoint{1.000000\du}{0\du}}{\pgfpoint{0\du}{1.000000\du}}
\pgfusepath{stroke}
\definecolor{dialinecolor}{rgb}{1.000000, 0.647059, 0.000000}
\pgfsetfillcolor{dialinecolor}
\pgfpathellipse{\pgfpoint{5.574152\du}{7.544911\du}}{\pgfpoint{1.000000\du}{0\du}}{\pgfpoint{0\du}{1.000000\du}}
\pgfusepath{fill}
\pgfsetlinewidth{0.100000\du}
\pgfsetdash{}{0pt}
\pgfsetdash{}{0pt}
\definecolor{dialinecolor}{rgb}{0.000000, 0.000000, 0.000000}
\pgfsetstrokecolor{dialinecolor}
\pgfpathellipse{\pgfpoint{5.574152\du}{7.544911\du}}{\pgfpoint{1.000000\du}{0\du}}{\pgfpoint{0\du}{1.000000\du}}
\pgfusepath{stroke}
\definecolor{dialinecolor}{rgb}{1.000000, 0.647059, 0.000000}
\pgfsetfillcolor{dialinecolor}
\pgfpathellipse{\pgfpoint{9.150804\du}{7.544911\du}}{\pgfpoint{1.000000\du}{0\du}}{\pgfpoint{0\du}{1.000000\du}}
\pgfusepath{fill}
\pgfsetlinewidth{0.100000\du}
\pgfsetdash{}{0pt}
\pgfsetdash{}{0pt}
\definecolor{dialinecolor}{rgb}{0.000000, 0.000000, 0.000000}
\pgfsetstrokecolor{dialinecolor}
\pgfpathellipse{\pgfpoint{9.150804\du}{7.544911\du}}{\pgfpoint{1.000000\du}{0\du}}{\pgfpoint{0\du}{1.000000\du}}
\pgfusepath{stroke}
\definecolor{dialinecolor}{rgb}{1.000000, 1.000000, 0.000000}
\pgfsetfillcolor{dialinecolor}
\pgfpathellipse{\pgfpoint{12.727456\du}{7.544911\du}}{\pgfpoint{1.000000\du}{0\du}}{\pgfpoint{0\du}{1.000000\du}}
\pgfusepath{fill}
\pgfsetlinewidth{0.100000\du}
\pgfsetdash{}{0pt}
\pgfsetdash{}{0pt}
\definecolor{dialinecolor}{rgb}{0.000000, 0.000000, 0.000000}
\pgfsetstrokecolor{dialinecolor}
\pgfpathellipse{\pgfpoint{12.727456\du}{7.544911\du}}{\pgfpoint{1.000000\du}{0\du}}{\pgfpoint{0\du}{1.000000\du}}
\pgfusepath{stroke}
\definecolor{dialinecolor}{rgb}{1.000000, 0.647059, 0.000000}
\pgfsetfillcolor{dialinecolor}
\pgfpathellipse{\pgfpoint{16.304108\du}{7.544911\du}}{\pgfpoint{1.000000\du}{0\du}}{\pgfpoint{0\du}{1.000000\du}}
\pgfusepath{fill}
\pgfsetlinewidth{0.100000\du}
\pgfsetdash{}{0pt}
\pgfsetdash{}{0pt}
\definecolor{dialinecolor}{rgb}{0.000000, 0.000000, 0.000000}
\pgfsetstrokecolor{dialinecolor}
\pgfpathellipse{\pgfpoint{16.304108\du}{7.544911\du}}{\pgfpoint{1.000000\du}{0\du}}{\pgfpoint{0\du}{1.000000\du}}
\pgfusepath{stroke}
\definecolor{dialinecolor}{rgb}{1.000000, 0.647059, 0.000000}
\pgfsetfillcolor{dialinecolor}
\pgfpathellipse{\pgfpoint{19.880759\du}{7.544911\du}}{\pgfpoint{1.000000\du}{0\du}}{\pgfpoint{0\du}{1.000000\du}}
\pgfusepath{fill}
\pgfsetlinewidth{0.100000\du}
\pgfsetdash{}{0pt}
\pgfsetdash{}{0pt}
\definecolor{dialinecolor}{rgb}{0.000000, 0.000000, 0.000000}
\pgfsetstrokecolor{dialinecolor}
\pgfpathellipse{\pgfpoint{19.880759\du}{7.544911\du}}{\pgfpoint{1.000000\du}{0\du}}{\pgfpoint{0\du}{1.000000\du}}
\pgfusepath{stroke}
\definecolor{dialinecolor}{rgb}{1.000000, 0.647059, 0.000000}
\pgfsetfillcolor{dialinecolor}
\pgfpathellipse{\pgfpoint{23.457411\du}{7.544911\du}}{\pgfpoint{1.000000\du}{0\du}}{\pgfpoint{0\du}{1.000000\du}}
\pgfusepath{fill}
\pgfsetlinewidth{0.100000\du}
\pgfsetdash{}{0pt}
\pgfsetdash{}{0pt}
\definecolor{dialinecolor}{rgb}{0.000000, 0.000000, 0.000000}
\pgfsetstrokecolor{dialinecolor}
\pgfpathellipse{\pgfpoint{23.457411\du}{7.544911\du}}{\pgfpoint{1.000000\du}{0\du}}{\pgfpoint{0\du}{1.000000\du}}
\pgfusepath{stroke}
\definecolor{dialinecolor}{rgb}{0.000000, 0.000000, 1.000000}
\pgfsetfillcolor{dialinecolor}
\pgfpathellipse{\pgfpoint{1.997500\du}{14.300000\du}}{\pgfpoint{1.000000\du}{0\du}}{\pgfpoint{0\du}{1.000000\du}}
\pgfusepath{fill}
\pgfsetlinewidth{0.100000\du}
\pgfsetdash{}{0pt}
\pgfsetdash{}{0pt}
\definecolor{dialinecolor}{rgb}{0.000000, 0.000000, 0.000000}
\pgfsetstrokecolor{dialinecolor}
\pgfpathellipse{\pgfpoint{1.997500\du}{14.300000\du}}{\pgfpoint{1.000000\du}{0\du}}{\pgfpoint{0\du}{1.000000\du}}
\pgfusepath{stroke}
\definecolor{dialinecolor}{rgb}{0.000000, 0.000000, 1.000000}
\pgfsetfillcolor{dialinecolor}
\pgfpathellipse{\pgfpoint{7.362478\du}{14.300000\du}}{\pgfpoint{1.000000\du}{0\du}}{\pgfpoint{0\du}{1.000000\du}}
\pgfusepath{fill}
\pgfsetlinewidth{0.100000\du}
\pgfsetdash{}{0pt}
\pgfsetdash{}{0pt}
\definecolor{dialinecolor}{rgb}{0.000000, 0.000000, 0.000000}
\pgfsetstrokecolor{dialinecolor}
\pgfpathellipse{\pgfpoint{7.362478\du}{14.300000\du}}{\pgfpoint{1.000000\du}{0\du}}{\pgfpoint{0\du}{1.000000\du}}
\pgfusepath{stroke}
\definecolor{dialinecolor}{rgb}{0.000000, 0.000000, 1.000000}
\pgfsetfillcolor{dialinecolor}
\pgfpathellipse{\pgfpoint{12.727456\du}{14.300000\du}}{\pgfpoint{1.000000\du}{0\du}}{\pgfpoint{0\du}{1.000000\du}}
\pgfusepath{fill}
\pgfsetlinewidth{0.100000\du}
\pgfsetdash{}{0pt}
\pgfsetdash{}{0pt}
\definecolor{dialinecolor}{rgb}{0.000000, 0.000000, 0.000000}
\pgfsetstrokecolor{dialinecolor}
\pgfpathellipse{\pgfpoint{12.727456\du}{14.300000\du}}{\pgfpoint{1.000000\du}{0\du}}{\pgfpoint{0\du}{1.000000\du}}
\pgfusepath{stroke}
\definecolor{dialinecolor}{rgb}{0.000000, 0.000000, 1.000000}
\pgfsetfillcolor{dialinecolor}
\pgfpathellipse{\pgfpoint{18.092433\du}{14.300000\du}}{\pgfpoint{1.000000\du}{0\du}}{\pgfpoint{0\du}{1.000000\du}}
\pgfusepath{fill}
\pgfsetlinewidth{0.100000\du}
\pgfsetdash{}{0pt}
\pgfsetdash{}{0pt}
\definecolor{dialinecolor}{rgb}{0.000000, 0.000000, 0.000000}
\pgfsetstrokecolor{dialinecolor}
\pgfpathellipse{\pgfpoint{18.092433\du}{14.300000\du}}{\pgfpoint{1.000000\du}{0\du}}{\pgfpoint{0\du}{1.000000\du}}
\pgfusepath{stroke}
\definecolor{dialinecolor}{rgb}{0.000000, 0.000000, 1.000000}
\pgfsetfillcolor{dialinecolor}
\pgfpathellipse{\pgfpoint{23.457411\du}{14.300000\du}}{\pgfpoint{1.000000\du}{0\du}}{\pgfpoint{0\du}{1.000000\du}}
\pgfusepath{fill}
\pgfsetlinewidth{0.100000\du}
\pgfsetdash{}{0pt}
\pgfsetdash{}{0pt}
\definecolor{dialinecolor}{rgb}{0.000000, 0.000000, 0.000000}
\pgfsetstrokecolor{dialinecolor}
\pgfpathellipse{\pgfpoint{23.457411\du}{14.300000\du}}{\pgfpoint{1.000000\du}{0\du}}{\pgfpoint{0\du}{1.000000\du}}
\pgfusepath{stroke}
\end{tikzpicture}